\documentclass[twocolumn]{revtex4}
\usepackage[utf8]{inputenc}
\usepackage{natbib}

\begin{document}

\title{Two-fluid hydrodynamics of cold atomic bosons under influence of the quantum fluctuations at non-zero temperatures}

\author{Pavel A. Andreev}
\email{andreevpa@physics.msu.ru}
\affiliation{Faculty of physics, Lomonosov Moscow State University, Moscow, Russian Federation, 119991.}
\affiliation{Peoples Friendship University of Russia (RUDN University), 6 Miklukho-Maklaya Street, Moscow, 117198, Russian Federation}

\date{\today}

\begin{abstract}
Ultracold Bose atoms is the physical system,
where the quantum and nonlinear phenomena play crucial role.
Ultracold bosons are considered at the small finite temperatures.
Bosons are considered as two different fluids: Bose-Einstein condensate and normal fluid (the thermal component).
An extended hydrodynamic model is obtained for both fluids,
where the pressure evolution equations and the pressure flux third rank tensor evolution equations
are considered along with the continuity and Euler equations.
It is found that the pressure evolution equation contains zero contribution of the short-range interaction.
The pressure flux evolution equation contains the interaction
which gives the quantum fluctuations in the zero temperature limit.
Here, we obtain its generalization for the finite temperature.
The contribution of interaction in the pressure flux evolution equation which goes to zero in the zero temperature limit is found.
The model is obtained via the straightforward derivation from the microscopic many-particle Schrodinger equation in the coordinate representation.
\end{abstract}



\maketitle


\section{Introduction}

Small temperature bosons are studied in terms of two-fluid hydrodynamics
consisting of the Bose-Einstein condensate (BEC) and normal fluid \cite{Dalfovo RMP 99}.
Each fluid is considered in terms of two hydrodynamic equations: the continuity and Euler equations.
It is assumed that
the BEC can be completely described by the concentration and velocity field,
or, in other terms, by the Gross-Pitaevskii equation \cite{Dalfovo RMP 99},
since BEC is the collection of particles in the single quantum state.
However, the normal fluid model requires an truncation of the set of hydrodynamic equations.
The pressure of normal fluid existing in the Euler equation for the normal fluid is an independent function.
Equation for the pressure evolution provides an expression for the pressure perturbations via the perturbations of other functions.
Application of the equation of state for pressure makes the model more simple,
but equation of state for pressure leads to the less accurate model.

Moreover, the kinetic pressure
in the Euler equation for the BEC is usually chosen to be equal to zero \cite{Andreev PRA08}, \cite{Andreev LP 19}.
Since the kinetic pressure is related to the occupation of the excited states
However, there is the nonzero part caused by the quantum fluctuations \cite{Andreev 2005}.
The pressure evolution equation of the weakly interacting bosons contains no interaction,
but next equation in the chain of the quantum hydrodynamic equations
(the equation for the pressure flux third rank tensor)
contains the interaction causing the depletion of the BECs at the zero temperature.
Therefore, it is necessary to consider the pressure flux evolution equation both for the BEC and for the normal fluid at the analysis of the small temperature influence.

The quantum depletion of the BECs is the appearance of the bosons in the excited states
while system is kept at the zero temperature.
So, some energy of the collective motion is transferred to the individual motion of a portion of particles.
It is caused by the quantum fluctuations related to the interparticle interaction.
The quantum fluctuations are considered in literature for a long time.
Mostly, their theoretical analysis is based on the Bogoliubov-de Gennes approach
\cite{Lee PR 57}, \cite{Pitaevskii PRL 98}, \cite{Braaten PRL 99}, \cite{Astrakharchik PRL 05}.
The quantum fluctuations in BECs are studied experimentally as well
\cite{Xu PRL 06}, \cite{Altmeyer PRL 07}, \cite{Papp PRL 08}.
This method is generalized for the dipolar BECs,
where the quantum fluctuations plays crucial role at the description of the dipolar BECs of lantanoids.
The dipolar lantanoid BECs reveal the large scale instability causing the splitting of the cloud of atoms
on the number of macroscopic drops.
This highly nonlinear phenomena is called the quantum droplet formation
\cite{Kadau Pfau Nature 16, Ferrier-Barbut PRL 16, Baillie PRA 16, Bisset PRA 16, Wachtler PRA 16 a1, Wachtler PRA 16 a2, Blakie pra 16, Boudjemaa PRA 20, Heinonen PRA 19, Malomed Phys D 19, Shamriz PRA 20, Li PRA 19, Aybar PRA 19, Examilioti JP B 20, Miyakawa PRA 20, Bottcher arXiv 20 07, Bisset arXiv 20 07, Wang arXiv 20 02, Edmonds arXiv 20 02, Baillie PRA 20}.
Therefore, reassemble of bosons in smaller compact groups causes the stabilization of the system.
These studies give the motivation for the study of quantum fluctuations.
However, here we restrict our analysis with the short-range interaction only and no dipole-dipole interaction is discussed.
Moreover, the influence of the finite temperature is the necessary part of complete model of these phenomena.

Fundamental feature of the collective dynamics in the spectrum of the sound waves.
Two distinct sound velocities exist in finite temperature ultracold Bose gas \cite{Dalfovo RMP 99}, \cite{Griffin PRB 96}.
The two-fluid model shows that the slower mode (second sound) is associated with the BEC component,
while the faster mode (first sound) is associated with the thermal component.
Generalized expressions for the speeds of sounds are obtained within developed model.

Derivation of two-fluid hydrodynamics for the finite temperature bosons in the limit of small temperature,
where the large fraction of the bosons is located in the BEC state is given from the microscopic motion in accordance with the quantum hydrodynamic method
\cite{Andreev PRA08}, \cite{Andreev LP 19}, \cite{MaksimovTMP 2001}, \cite{Andreev PTEP 19}.
The microscopic dynamics is described by the Schrodinger equation in the coordinate representation.
Collection of the macroscopic functions is presented
to describe the collective effects in ultracold bosons.
The last includes the concentration of particles, the velocity field and the pressure tensor.
The derivation of basic equations is made for all bosons distributed on the lower energy level and the excited levels as the single fluid.
The decomposition on two fluids is made on the microscopic scale.
After general structure of equations is obtained for the arbitrary temperature and arbitrary strength of interaction,
an approximate calculation of functions presenting the interaction is made for the regime of short-range interaction.
Hence, the small parameter related to the small area of interaction potential is used.
It gives a specification for general model,
but also the first order contribution on the small parameter is applied.
The further truncation is made at calculation of the interaction terms for weak interaction and small temperatures.

This paper is organized as follows.
In Sec. II major steps of derivation of hydrodynamic equations from the Schrodinger equation are demonstrated,
where the pressure evolution equation (the quantum Bohm potential evolution equation) and the third rank tensor evolution equation are obtained
along with the continuity and Euler equations.
In Sec. III calculation of interaction in the Euler equation, the pressure evolution equation, and the pressure flux evolution equation is demonstrated.
In Sec. IV presents the suggested version of the extended two-fluid quantum hydrodynamic model for the ultracold finite temperature bosons.
In Sec. V the limiting regime of derived model for the BEC is obtained under influence of the quantum fluctuations at the zero temperature.
In Sec. VI a brief summary of obtained results is presented.

\section{On derivation of hydrodynamic equations from microscopic quantum dynamics}

\subsection{Basic definitions of quantum hydrodynamics and the Euler equation derivation}

On the microscopic level we do not have notion of temperature.
Hence we consider system of interacting bosons governed by the Schrodinger equation $\imath\hbar\partial_{t}\Psi=\hat{H}\Psi$
with the following Hamiltonian
\begin{equation}\label{BECTP20 Hamiltonian micro}
\hat{H}=\sum_{i=1}^{N}\biggl(\frac{\hat{\textbf{p}}^{2}_{i}}{2m_{i}}+V_{ext}(\textbf{r}_{i},t)\biggr)
+\frac{1}{2}\sum_{i,j\neq i}U(\textbf{r}_{i}-\textbf{r}_{j}) ,\end{equation}
where $m_{i}$ is the mass of i-th particle,
$\hat{\textbf{p}}_{i}=-\imath\hbar\nabla_{i}$ is the momentum of i-th particle.
The last term in the Hamiltonian (\ref{BECTP20 Hamiltonian micro})
is the boson-boson interaction $U_{ij}$.
We do not specify the form of interaction.
However, the derivation presented below employs
that the interaction has finite value on the small distances between particles
and shows fast decay at the increase of the interparticle distance.
Definitely, no distinguishing between bosons in the BEC state and bosons in other states is made at this stage.
Separation of all bosons on two subsystems is made in terms of collective variables.

Hydrodynamic model usually made for each species of particles.
If we consider a single species then all masses equal to each other.

Distribution of particles in a trap, waves, solitons, oscillations of form of trapped particles are described by the concentration of particles.
Concentration is an essential macroscopic function both for classical and for quantum fluids.
The module of the macroscopic wave function in the Gross-Pitaevskii equation gives
the square root of concentration of bosons in the BEC state.
Therefore, we start the derivation of quantum hydrodynamic equations from the definition of concentration.
The quantum mechanics is based on notion of point-like objects
in spite the wave nature of quantum objects.
So, the eigenfunction of the coordinate operator in the coordinate representation is the delta function
$\hat{x}\psi_{x'}(x)=x'\psi_{x'}(x)$,
where normalized wave function is $\psi_{x'}(x)=\delta(x-x')$.
Obviously, the operation of concentration in the coordinate representation of quantum mechanics is the sum of delta functions
$\hat{n}=\sum_{i=1}^{N}\delta(\textbf{r}-\textbf{r}_{i})$.
Moreover, it is supported by the general rule for quantization.
We need to take corresponding classical function.

Transition to description of the collective motion of bosons is made via introduction of the concentration
\cite{Andreev PRA08}, \cite{MaksimovTMP 2001}:
\begin{equation}\label{BECTP20 concentration def b} n=\int
dR\sum_{i=1}^{N}\delta(\textbf{r}-\textbf{r}_{i})\Psi^{*}(R,t)\Psi(R,t),\end{equation}
which is the first collective variable in our model.
Other collective variables appear during the derivation.
Equation (\ref{BECTP20 concentration def b}) contains the following notations
$dR=\prod_{i=1}^{N}d\textbf{r}_{i}$ is the element of volume in $3N$ dimensional configurational space,
with $N$ is the number of bosons.
Concentration (\ref{BECTP20 concentration def b}) is the sum of partial concentrations $n=n_{n}+n_{b}$ describing
the distribution of BEC $n_{b}$ and normal fluid $n_{n}$ in the coordinate space.

The equation for evolution of concentration (\ref{BECTP20 concentration def b}) can be obtained by
acting by time derivative on function (\ref{BECTP20 concentration def b}).
The time derivative acts on the wave functions under the integral
while the time derivatives of the wave function are taken from the Schrodinger equation.
Obtain the continuity equation for concentration (\ref{BECTP20 concentration def b}) after straightforward calculations
\begin{equation}\label{BECTP20 cont eq via j} \partial_{t}n+\nabla\cdot \textbf{j}=0, \end{equation}
where the new collective function called the current appears as the following integral of the wave function
$$\textbf{j}(\textbf{r},t)
=\int dR\sum_{i=1}^{N}\delta(\textbf{r}-\textbf{r}_{i})\times$$
\begin{equation}\label{BECTP20 j def}
\times\frac{1}{2m_{i}}(\Psi^{*}(R,t)\hat{\textbf{p}}_{i}\Psi(R,t)+c.c.),\end{equation}
with $c.c.$ is the complex conjugation.

Both introduced collective functions $n(\textbf{r},t)$ and $\textbf{j}(\textbf{r},t)$ are quadratic forms of the wave function.
Each of them can be splitted on two parts related to the BEC and normal fluid.
Hence we have $n=n_{n}+n_{b}$ and $\textbf{j}=\textbf{j}_{n}+\textbf{j}_{b}$.
No microscopic definitions are introduced for the partial functions
$n_{n}$, $n_{b}$, $\textbf{j}_{n}$, and $\textbf{j}_{b}$.
Therefore, the continuity equation (\ref{BECTP20 cont eq via j}) splits on two partial continuity equations
\begin{equation}\label{BECTP20 cont eq via j for subspecies} \partial_{t}n_{a}+\nabla\cdot \textbf{j}_{a}=0, \end{equation}
where subindex $a$ stands for $b$ and $n$.

Continue the derivation of hydrodynamic equations and consider the time evolution of the particle current (\ref{BECTP20 j def}).
Act by time derivative on function $\textbf{j}$ (\ref{BECTP20 j def}) and use the Schrodinger equation with Hamiltonian (\ref{BECTP20 Hamiltonian micro}).
It leads to the general form of the current evolution equation
\begin{equation} \label{BECTP20 Euler eq 1 via j}
\partial_{t}j^{\alpha}+\partial_{\beta}\Pi^{\alpha\beta}
=-\frac{1}{m}n\partial_{\alpha}V_{ext}+\frac{1}{m}F^{\alpha}_{int}, \end{equation}
where
$$\Pi^{\alpha\beta}=\int dR\sum_{i=1}^{N}\delta(\textbf{r}-\textbf{r}_{i}) \frac{1}{4m^{2}}
[\Psi^{*}(R,t)\hat{p}_{i}^{\alpha}\hat{p}_{i}^{\beta}\Psi(R,t)$$
\begin{equation} \label{BECTP20 Pi def} +\hat{p}_{i}^{\alpha *}\Psi^{*}(R,t)\hat{p}_{i}^{\beta}\Psi(R,t)+c.c.] \end{equation}
is the momentum flux,
and
\begin{equation} \label{BECTP20 F alpha def via n2}
F^{\alpha}_{int}=-\int (\partial^{\alpha}U(\textbf{r}-\textbf{r}'))
n_{2}(\textbf{r},\textbf{r}',t)d\textbf{r}', \end{equation}
with the two-particle concentration
$$n_{2}(\textbf{r},\textbf{r}',t)$$
\begin{equation} \label{BECTP20 n2 def} =\int
dR\sum_{i,j=1,j\neq i}^{N}\delta(\textbf{r}-\textbf{r}_{i})\delta(\textbf{r}'-\textbf{r}_{j})\Psi^{*}(R,t)\Psi(R,t) .\end{equation}

It is necessary to split equation (\ref{BECTP20 Euler eq 1 via j}) on two equations for each subsystem of bosons.
In current form equation (\ref{BECTP20 Euler eq 1 via j}) consist of superposition of functions
which are quadratic forms of the wave function.
Hence, each term can be splitted on two parts and we find two similar equations for the currents
\begin{equation} \label{BECTP20 Euler eq 1 via j for subspecies}
\partial_{t}j_{a}^{\alpha}+\partial_{\beta}\Pi_{a}^{\alpha\beta}
=-\frac{1}{m}n_{a}\partial_{\alpha}V_{ext}+\frac{1}{m}F^{\alpha}_{a,int}. \end{equation}
The first and third terms are proportional to the concentration and the current.
therefore, they require no comments.
Nontrivial difference between two current evolution equation appears
at further analysis of the momentum flux $\Pi^{\alpha\beta}$ and the interaction $F^{\alpha}_{int}$.
However, we point out some difference which appear for the momentum flux $\Pi^{\alpha\beta}$.
Its structure is obtained in many papers
(see for instance \cite{Andreev PRA08} after equation (52), \cite{Andreev 2001} equation (24))
\begin{equation} \label{BECTP20 Pi via n v p T}\Pi^{\alpha\beta}=nv^{\alpha}v^{\beta} +p^{\alpha\beta}+T^{\alpha\beta},\end{equation}
where $p^{\alpha\beta}$ is the pressure tensor, and $T^{\alpha\beta}$ is the tensor giving the quantum Bohm potential,
its approximate form can be written in the following form
\begin{equation} \label{BECTP20 Bohm tensor single part}
T^{\alpha\beta}_{0}=-\frac{\hbar^{2}}{4m^{2}}\biggl[\partial_{\alpha}\partial_{\beta}n
-\frac{\partial_{\alpha}n\cdot\partial_{\beta}n}{n}\biggr].\end{equation}
Tensor $T^{\alpha\beta}_{0}$ (\ref{BECTP20 Bohm tensor single part}) is obtained for noninteracting particles located in the single quantum state.

Basically, the pressure tensor $p^{\alpha\beta}$ is  defined via the wave function $\Psi(R,t)$.
However, it requires some manipulations with the wave function and introduction of a number of intermediate function.
Hence we do not present its explicit form.
Nevertheless, the pressure tensor is related to the distribution of bosons on quantum states with energies above $E_{min}$.
Therefore, for bosons in the BEC state we have $p^{\alpha\beta}_{B}=0$ if no quantum fluctuations are considered and
\begin{equation} \label{BECTP20 Pi via n v p T BEC}\Pi_{B}^{\alpha\beta}=n_{B}v_{B}^{\alpha}v_{B}^{\beta} +T_{B}^{\alpha\beta}+p_{qf}^{\alpha\beta},\end{equation}
where $T_{B}^{\alpha\beta}$ is the function of $n_{B}$ (\ref{BECTP20 Bohm tensor single part}) if there is no interaction.
Distribution of particles on different quantum states does not allow to get full expression (\ref{BECTP20 Bohm tensor single part}),
but the first term.
However, it can be used as an equation of state for noninteracting limit.
The normal fluid bosons have nonzero pressure $p_{n}^{\alpha\beta}\neq0$.
Hence, all terms in presentation (\ref{BECTP20 Pi via n v p T}) exists in this regime.
Expression (\ref{BECTP20 Bohm tensor single part}) appears for bosons in the single state in the absence of interaction.
Hence, it is an approximate expression for the weakly interacting bosons being in the BEC state.
It is even less accurate for normal fluid bosons,
but we use it as an equation of state for the quantum part of the momentum flux.

\subsection{The pressure evolution equation}

Extending the set of hydrodynamic equations we can derive the equation for the momentum flux evolution.
It can be expected that
this equation brings extra information for the normal fluid bosons only.
However, the quantum fluctuations give contribution in the evolution of the kinetic pressure of BECs in the limit of zero temperature
via the divergence of the third rank tensor.
If the temperature is nonzero
we have two partial kinetic pressures for the BEC and for the normal fluid.
Consider the time evolution of the momentum flux (\ref{BECTP20 Pi def}) using the Schrodinger equation with Hamiltonian (\ref{BECTP20 Hamiltonian micro}).

It gives to the following expression:
$$\partial_{t}\Pi^{\alpha\beta}=\frac{\imath}{\hbar}\int dR\sum_{i=1}^{N}\delta(\textbf{r}-\textbf{r}_{i}) \frac{1}{4m^{2}}
[\hat{H}^{*}\Psi^{*}(R,t)\hat{p}_{i}^{\alpha}\hat{p}_{i}^{\beta}\Psi(R,t)$$
$$-\Psi^{*}(R,t)\hat{p}_{i}^{\alpha}\hat{p}_{i}^{\beta}\hat{H}\Psi(R,t)
+\hat{p}_{i}^{\alpha *}\hat{H}^{*}\Psi^{*}(R,t)\hat{p}_{i}^{\beta}\Psi(R,t)$$
\begin{equation} \label{BECTP20 Pi time deriv 1}
-\hat{p}_{i}^{\alpha *}\Psi^{*}(R,t)\hat{p}_{i}^{\beta}\hat{H}\Psi(R,t)-c.c.] \end{equation}
The part of the presented terms contains the Hamiltonian $\hat{H}$ under action of the momentum operators.
We permute the Hamiltonian $\hat{H}$ and the operators acting on it.
Hence, the result of permutation presented by terms
where no operators act on the Hamiltonian $\hat{H}$.
However, the terms containing the corresponding commutators appear.
Therefore, all terms are combined in two groups:
$$\partial_{t}\Pi^{\alpha\beta}=\frac{\imath}{\hbar}\int dR\sum_{i=1}^{N}\delta(\textbf{r}-\textbf{r}_{i}) \frac{1}{4m^{2}}
[\hat{H}^{*}\Psi^{*}\cdot\hat{p}_{i}^{\alpha}\hat{p}_{i}^{\beta}\Psi$$
$$-\Psi^{*}\hat{H}\hat{p}_{i}^{\alpha}\hat{p}_{i}^{\beta}\Psi
+\hat{H}^{*}\hat{p}_{i}^{\alpha *}\Psi^{*}\cdot\hat{p}_{i}^{\beta}\Psi
-\hat{p}_{i}^{\alpha *}\Psi^{*}\cdot\hat{H}\hat{p}_{i}^{\beta}\Psi-c.c.] $$
$$+\frac{\imath}{\hbar}\int dR\sum_{i=1}^{N}\delta(\textbf{r}-\textbf{r}_{i}) \frac{1}{4m^{2}}
[-\Psi^{*}[\hat{p}_{i}^{\alpha}\hat{p}_{i}^{\beta},\hat{H}]\Psi$$
\begin{equation} \label{BECTP20 Pi time deriv 2} +[\hat{p}_{i}^{\alpha *},\hat{H}^{*}]\Psi^{*}\cdot\hat{p}_{i}^{\beta}\Psi
-\hat{p}_{i}^{\alpha *}\Psi^{*}\cdot[\hat{p}_{i}^{\beta},\hat{H}]\Psi-c.c.] \end{equation}
The first group of terms in expression (\ref{BECTP20 Pi time deriv 2}) gives the divergence of flux of tensor $\Pi^{\alpha\beta}$.
The second group of terms contains the commutators.
This group leads to the contribution of interaction in the momentum flux evolution.

It gives the momentum flux evolution equation
$$\partial_{t}\Pi^{\alpha\beta}+\partial_{\gamma}M^{\alpha\beta\gamma}
=-\frac{1}{m}j^{\beta}\partial_{\alpha}V_{ext}$$
\begin{equation} \label{BECTP20 eq for Pi alpha beta} -\frac{1}{m}j^{\alpha}\partial_{\beta}V_{ext}
+\frac{1}{m}(F^{\alpha\beta}+F^{\beta\alpha}), \end{equation}
where the momentum flux is the full flux of all bosons
$\Pi^{\alpha\beta}=\Pi_{n}^{\alpha\beta}+\Pi_{b}^{\alpha\beta}$,
the splitting on two subspecies is to be made later,
\begin{equation} \label{BECTP20 F alpha beta def} F^{\alpha\beta}=-\int[\partial^{\alpha}U(\textbf{r}-\textbf{r}')]j_{2}^{\beta}(\textbf{r},\textbf{r}',t)d\textbf{r}'\end{equation}
is the force tensor field,
$$M^{\alpha\beta\gamma}=\int dR\sum_{i=1}^{N}\delta(\textbf{r}-\textbf{r}_{i}) \frac{1}{8m_{i}^{3}}\biggl[\Psi^{*}(R,t)\hat{p}_{i}^{\alpha}\hat{p}_{i}^{\beta}\hat{p}_{i}^{\gamma}\Psi(R,t)$$
$$+\hat{p}_{i}^{\alpha *}\Psi^{*}(R,t)\hat{p}_{i}^{\beta}\hat{p}_{i}^{\gamma}\Psi(R,t)
+\hat{p}_{i}^{\alpha *}\hat{p}_{i}^{\gamma *}\Psi^{*}(R,t)\hat{p}_{i}^{\beta}\Psi(R,t)$$
\begin{equation} \label{BECTP20 M alpha beta gamma def}
+\hat{p}_{i}^{\gamma *}\Psi^{*}(R,t)\hat{p}_{i}^{\alpha}\hat{p}_{i}^{\beta}\Psi(R,t)+c.c.\biggr] \end{equation}
is the current (flux) of the momentum flux,
and
$$\textbf{j}_{2}(\textbf{r},\textbf{r}',t)=\int
dR\sum_{i,j\neq i}\delta(\textbf{r}-\textbf{r}_{i})\delta(\textbf{r}'-\textbf{r}_{j})\times$$
\begin{equation} \label{BECTP20 j 2 def}
\times\frac{1}{2m_{i}}[\Psi^{*}(R,t)\hat{\textbf{p}}_{i}\Psi(R,t)+c.c.] \end{equation}
is the two-particle current-concentration function.

If quantum correlations are dropped function $j_{2}^{\alpha}(\textbf{r},\textbf{r}',t)$
splits on product of the current $j^{\alpha}(\textbf{r},t)$ and the concentration $n(\textbf{r}',t)$.
Interaction in the momentum flux evolution equation (\ref{BECTP20 eq for Pi alpha beta}) is presented
by symmetrized combinations of tensors $F^{\alpha\beta}$,
which is the flux or current of force.

Partial momentum flux equations appear as
$$\partial_{t}\Pi_{a}^{\alpha\beta}+\partial_{\gamma}M_{a}^{\alpha\beta\gamma}
=-\frac{1}{m}j_{a}^{\beta}\partial_{\alpha}V_{ext}$$
\begin{equation} \label{BECTP20 eq for Pi alpha beta Partial} -\frac{1}{m}j_{a}^{\alpha}\partial_{\beta}V_{ext}
+\frac{1}{m}(F_{a}^{\alpha\beta}+F_{a}^{\beta\alpha}), \end{equation}
where $M^{\alpha\beta\gamma}=M_{B}^{\alpha\beta\gamma}+M_{n}^{\alpha\beta\gamma}$,
with
$$M_{a}^{\alpha\beta\gamma}= n_{a}v_{a}^{\alpha}v_{a}^{\beta}v_{a}^{\gamma}
+v_{a}^{\alpha} (p_{a}^{\beta\gamma}+T_{a}^{\beta\gamma})
+v_{a}^{\beta} (p_{a}^{\alpha\gamma}+T_{a}^{\alpha\gamma}) $$
\begin{equation} \label{BECTP20 M via p T Q}
+v_{a}^{\gamma} (p_{a}^{\alpha\beta}+T_{a}^{\alpha\beta})
+Q_{a}^{\alpha\beta\gamma}+T_{a}^{\alpha\beta\gamma}
+L_{a}^{\alpha\beta\gamma} . \end{equation}
The pressure is the average of the square of the thermal velocity,
when tensor $Q_{a}^{\alpha\beta\gamma}$ is the average of the product of three projections of the thermal velocity.
Function $L_{a}^{\alpha\beta\gamma}$ presents quantum-thermal terms.
For the BEC we have $p_{B}^{\alpha\beta}=0$, $Q_{B}^{\alpha\beta\gamma}=0$, $L_{B}^{\alpha\beta\gamma}=0 $,
since it has no contribution of the excited states.
For symmetric equilibrium distributions we have $Q_{n}^{\alpha\beta\gamma}=0$, $L_{n}^{\alpha\beta\gamma}=0$.
We generalize this conclusion for nonequilibrium states as the trivial equations of state for these functions.
Tensor $T_{a}^{\alpha\beta\gamma}$ is
\begin{equation} \label{BECTP20 T abc}T_{a}^{\alpha\beta\gamma}
=-\frac{\hbar^{2}}{12m^{2}}n_{a}(\partial^{\alpha}\partial^{\beta} v_{a}^{\gamma}
+\partial^{\alpha}\partial^{\gamma} v_{a}^{\beta}
+\partial^{\beta}\partial^{\gamma} v_{a}^{\alpha}).\end{equation}
This definition of tensor $T^{\alpha\beta\gamma}$ differs from equation (27) in Ref. \cite{Andreev 2001}
by extraction of the quantum Bohm potentials written together with pressure tensors in equation (\ref{BECTP20 M via p T Q}).
Equation (27) in Ref. \cite{Andreev 2001} contains approximate form of the quantum Bohm potential $T^{\alpha\beta}$.
Equation (\ref{BECTP20 M via p T Q}) includes the quantum Bohm potential in its general form.
Moreover, expression (\ref{BECTP20 T abc}) is an exact formula obtained with no assumption about structure of the many-particle wave function
like the first term in equation (23) in Ref. \cite{Andreev 2001}.

Equations (\ref{BECTP20 cont eq via j})-(\ref{BECTP20 eq for Pi alpha beta Partial}) are obtained in general form.
The short-range nature of the inter-particle interaction is not used.
Moreover, the traditional hydrodynamic equations are presented via the velocity field and the pressure tensor
while equations (\ref{BECTP20 cont eq via j})-(\ref{BECTP20 eq for Pi alpha beta Partial}) are written via the current and the momentum flux.

The method of the introduction of the velocity field in the equations of quantum hydrodynamics of spinless particles
is presented in Refs. \cite{Andreev PRA08}, \cite{Andreev 2001}.
The method of calculation of the terms containing interaction for the short-range interaction limit
is also described in Refs. \cite{Andreev PRA08}, \cite{Andreev 2001}.
Let us present results of application of these methods for finite temperature bosons.
Moreover, we consider the short-range interaction in the first order by the interaction radius.

\subsection{Appearance of the quantum fluctuations in the third rank tensor evolution equation}

Derivation of the quantum fluctuations requires the calculation of the time evolution of the current of the momentum flux
$M^{\alpha\beta\gamma}$ (\ref{BECTP20 M alpha beta gamma def}).
The method of derivation is similar to the equations obtained above.
The time derivative of tensor $M^{\alpha\beta\gamma}$ acts on the wave function in its definition.
The time derivative of the wave function is replaced by the Hamiltonian (\ref{BECTP20 Hamiltonian micro})
in accordance with the many-particle microscopic Schrodinger equation $\imath\hbar\partial_{t}\Psi=\hat{H}\Psi$.
It leads to the following expression:
$$\partial_{t}M^{\alpha\beta\gamma}=\frac{\imath}{\hbar}\int dR\sum_{i=1}^{N}\delta(\textbf{r}-\textbf{r}_{i}) \frac{1}{8m_{i}^{3}}\biggl[\hat{H}^{*}\Psi^{*}\cdot\hat{p}_{i}^{\alpha}\hat{p}_{i}^{\beta}\hat{p}_{i}^{\gamma}\Psi$$
$$-\Psi^{*}\hat{p}_{i}^{\alpha}\hat{p}_{i}^{\beta}\hat{p}_{i}^{\gamma}\hat{H}\Psi
+\hat{p}_{i}^{\alpha *}\hat{H}^{*}\Psi^{*}\cdot\hat{p}_{i}^{\beta}\hat{p}_{i}^{\gamma}\Psi
+\hat{p}_{i}^{\alpha *}\hat{p}_{i}^{\gamma *}\hat{H}^{*}\Psi^{*}\cdot\hat{p}_{i}^{\beta}\Psi$$
$$-\hat{p}_{i}^{\alpha *}\Psi^{*}\cdot\hat{p}_{i}^{\beta}\hat{p}_{i}^{\gamma}\hat{H}\Psi
-\hat{p}_{i}^{\alpha *}\hat{p}_{i}^{\gamma *}\Psi^{*}\cdot\hat{p}_{i}^{\beta}\hat{H}\Psi$$
\begin{equation} \label{BECTP20 M alpha beta gamma time deriv 1}
+\hat{p}_{i}^{\gamma *}\hat{H}^{*}\Psi^{*}\cdot\hat{p}_{i}^{\alpha}\hat{p}_{i}^{\beta}\Psi
-\hat{p}_{i}^{\gamma *}\Psi^{*}\cdot\hat{p}_{i}^{\alpha}\hat{p}_{i}^{\beta}\hat{H}\Psi-c.c.\biggr] \end{equation}
The part of the presented terms contains the Hamiltonian $\hat{H}$ under action of the momentum operators.
We permute the Hamiltonian $\hat{H}$ and the operators acting on it.
Hence, the result of permutation presented by terms
where no operators act on the Hamiltonian $\hat{H}$.
However, the terms containing the corresponding commutators appear.
Therefore, all terms are combined in two groups:
$$\partial_{t}M^{\alpha\beta\gamma}=\frac{\imath}{\hbar}\int dR\sum_{i=1}^{N}\delta(\textbf{r}-\textbf{r}_{i}) \frac{1}{8m_{i}^{3}}\biggl[\hat{H}^{*}\Psi^{*}\cdot\hat{p}_{i}^{\alpha}\hat{p}_{i}^{\beta}\hat{p}_{i}^{\gamma}\Psi$$
$$-\Psi^{*}\hat{H}\hat{p}_{i}^{\alpha}\hat{p}_{i}^{\beta}\hat{p}_{i}^{\gamma}\Psi
+\hat{H}^{*}\hat{p}_{i}^{\alpha *}\Psi^{*}\cdot\hat{p}_{i}^{\beta}\hat{p}_{i}^{\gamma}\Psi
+\hat{H}^{*}\hat{p}_{i}^{\alpha *}\hat{p}_{i}^{\gamma *}\Psi^{*}\cdot\hat{p}_{i}^{\beta}\Psi$$
$$-\hat{p}_{i}^{\alpha *}\Psi^{*}\cdot\hat{H}\hat{p}_{i}^{\beta}\hat{p}_{i}^{\gamma}\Psi
-\hat{p}_{i}^{\alpha *}\hat{p}_{i}^{\gamma *}\Psi^{*}\cdot\hat{H}\hat{p}_{i}^{\beta}\Psi$$
$$+\hat{H}^{*}\hat{p}_{i}^{\gamma *}\Psi^{*}\cdot\hat{p}_{i}^{\alpha}\hat{p}_{i}^{\beta}\Psi
-\hat{p}_{i}^{\gamma *}\Psi^{*}\cdot\hat{H}\hat{p}_{i}^{\alpha}\hat{p}_{i}^{\beta}\Psi-c.c.\biggr]$$

$$+\frac{\imath}{\hbar}\int dR\sum_{i=1}^{N}\delta(\textbf{r}-\textbf{r}_{i})
\frac{1}{8m_{i}^{3}}\biggl[
-\Psi^{*}[\hat{p}_{i}^{\alpha}\hat{p}_{i}^{\beta}\hat{p}_{i}^{\gamma},\hat{H}]\Psi$$
$$+[\hat{p}_{i}^{\alpha *},\hat{H}^{*}]\Psi^{*}\cdot\hat{p}_{i}^{\beta}\hat{p}_{i}^{\gamma}\Psi
+[\hat{p}_{i}^{\alpha *}\hat{p}_{i}^{\gamma *},\hat{H}^{*}]\Psi^{*}\cdot\hat{p}_{i}^{\beta}\Psi$$
$$-\hat{p}_{i}^{\alpha *}\Psi^{*}\cdot[\hat{p}_{i}^{\beta}\hat{p}_{i}^{\gamma},\hat{H}]\Psi
-\hat{p}_{i}^{\alpha *}\hat{p}_{i}^{\gamma *}\Psi^{*}\cdot[\hat{p}_{i}^{\beta},\hat{H}]\Psi$$
\begin{equation} \label{BECTP20 M alpha beta gamma time deriv 2}
+[\hat{p}_{i}^{\gamma *},\hat{H}^{*}]\Psi^{*}\cdot\hat{p}_{i}^{\alpha}\hat{p}_{i}^{\beta}\Psi
-\hat{p}_{i}^{\gamma *}\Psi^{*}\cdot[\hat{p}_{i}^{\alpha}\hat{p}_{i}^{\beta},\hat{H}]\Psi-c.c.\biggr].
\end{equation}
The first group of terms leads to the divergence of the flux of tensor $M^{\alpha\beta\gamma}$.
The second group of terms containing the commutators presents the interactions.

Final form of tensor $M^{\alpha\beta\gamma}$ evolution equation can be expressed in the following terms:
$$\partial_{t}M^{\alpha\beta\gamma}+\partial_{\delta}R^{\alpha\beta\gamma\delta}
=\frac{\hbar^{2}}{4m^{3}}n\partial_{\alpha}\partial_{\beta}\partial_{\gamma}V_{ext} $$
$$-\frac{1}{m}\Pi^{\beta\gamma}\partial_{\alpha}V_{ext}
-\frac{1}{m}\Pi^{\alpha\gamma}\partial_{\beta}V_{ext}
-\frac{1}{m}\Pi^{\alpha\beta}\partial_{\gamma}V_{ext} $$
\begin{equation} \label{BECTP20 eq for M alpha beta gamma}
+\frac{1}{m}F_{qf}^{\alpha\beta\gamma}
+\frac{1}{m}(F^{\alpha\beta\gamma}+F^{\beta\gamma\alpha}+F^{\gamma\alpha\beta}), \end{equation}
where
\begin{equation} \label{BECTP20 F alpha beta gamma qf def} F_{qf}^{\alpha\beta\gamma}=\frac{\hbar^{2}}{4m^{2}} \int[\partial^{\alpha}\partial^{\beta}\partial^{\gamma}U(\textbf{r}-\textbf{r}')]n_{2}(\textbf{r},\textbf{r}',t)d\textbf{r}'
\end{equation}
is the quantum force contribution leading to the quantum fluctuations,
and
\begin{equation} \label{BECTP20 F alpha beta gamma def} F^{\alpha\beta\gamma}=-\int[\partial^{\alpha}U(\textbf{r}-\textbf{r}')]\Pi_{2}^{\beta\gamma}(\textbf{r},\textbf{r}',t)d\textbf{r}'
\end{equation}
is the interaction contribution containing nonzero limit in the classical regime,
with
$$\Pi_{2}^{\alpha\beta}(\textbf{r},\textbf{r}',t)=\int
dR\sum_{i,j\neq i}\frac{1}{4m_{i}^{2}}\delta(\textbf{r}-\textbf{r}_{i})\delta(\textbf{r}'-\textbf{r}_{j})\times$$
\begin{equation} \label{BECTP20 Pi 2 def}
\times[\Psi^{*}\hat{p}_{i}^{\alpha}\hat{p}_{i}^{\beta}\Psi
+(\hat{p}_{i}^{\beta})^{*}\Psi^{*}\hat{p}_{i}^{\alpha}\Psi+c.c.]. \end{equation}
Tensor $\Pi_{2}^{\alpha\beta}(\textbf{r},\textbf{r}',t)$ can be simplified in the correlationless regime
to the following form $\Pi_{2}^{\alpha\beta}(\textbf{r},\textbf{r}',t)=\Pi^{\alpha\beta}(\textbf{r},t)\cdot n(\textbf{r}',t)$.
However, the correlations caused by the symmetrization of the bosonic many-particle wave function are used below.

Terms $F^{\alpha\beta\gamma}$ and $F_{qf}^{\alpha\beta\gamma}$ are
the third rank force tensors describing the interparticle interaction.
However, equation (\ref{BECTP20 eq for M alpha beta gamma}) contains the flux of tensor $M^{\alpha\beta\gamma}$
which is the fourth rank tensor appearing in the following form:
$$R^{\alpha\beta\gamma\delta}=\int dR\sum_{i=1}^{N}\delta(\textbf{r}-\textbf{r}_{i}) \frac{1}{16m_{i}^{4}}\biggl[\Psi^{*}\hat{p}_{i}^{\alpha}\hat{p}_{i}^{\beta}\hat{p}_{i}^{\gamma}\hat{p}_{i}^{\delta}\Psi$$
$$+\hat{p}_{i}^{\alpha *}\Psi^{*}\hat{p}_{i}^{\beta}\hat{p}_{i}^{\gamma}\hat{p}_{i}^{\delta}\Psi
+\hat{p}_{i}^{\beta *}\Psi^{*}\hat{p}_{i}^{\alpha}\hat{p}_{i}^{\gamma}\hat{p}_{i}^{\delta}\Psi
+\hat{p}_{i}^{\gamma *}\Psi^{*}\hat{p}_{i}^{\alpha}\hat{p}_{i}^{\beta}\hat{p}_{i}^{\gamma}\Psi$$
$$+\hat{p}_{i}^{\delta *}\Psi^{*}\hat{p}_{i}^{\alpha}\hat{p}_{i}^{\beta}\hat{p}_{i}^{\gamma}\Psi
+\hat{p}_{i}^{\alpha *}\hat{p}_{i}^{\delta *}\Psi^{*}\hat{p}_{i}^{\beta}\hat{p}_{i}^{\gamma}\Psi$$
\begin{equation} \label{BECTP20 R alpha beta gamma delta def}
+\hat{p}_{i}^{\alpha *}\hat{p}_{i}^{\gamma *}\Psi^{*}\hat{p}_{i}^{\beta}\hat{p}_{i}^{\delta}\Psi
+\hat{p}_{i}^{\gamma *}\hat{p}_{i}^{\delta *}\Psi^{*}\hat{p}_{i}^{\alpha}\hat{p}_{i}^{\beta}\Psi+c.c.\biggr]. \end{equation}

Equation (\ref{BECTP20 eq for M alpha beta gamma}) is obtained for bosons with the arbitrary temperature.
It can be separated on two equations for two following subsystems:
the BEC and the normal fluid.
All terms in equation (\ref{BECTP20 eq for M alpha beta gamma}) are additive on the particles.
Therefore, they are additive on the subsystems.
Hence, the structure of the partial equations is identical to the structure of equation (\ref{BECTP20 eq for M alpha beta gamma}):
$$\partial_{t}M_{a}^{\alpha\beta\gamma}+\partial_{\delta}R_{a}^{\alpha\beta\gamma\delta}=
-\frac{1}{m}n_{a}\partial_{\alpha}\partial_{\beta}\partial_{\gamma}V_{ext} $$
$$-\frac{1}{m}\Pi_{a}^{\beta\gamma}\partial_{\alpha}V_{ext}
-\frac{1}{m}\Pi_{a}^{\alpha\gamma}\partial_{\beta}V_{ext}
-\frac{1}{m}\Pi_{a}^{\alpha\beta}\partial_{\gamma}V_{ext} $$
\begin{equation} \label{BECTP20 eq for M alpha beta gamma partial}
+\frac{1}{m}F_{a,qf}^{\alpha\beta\gamma}
+\frac{1}{m}(F_{a}^{\alpha\beta\gamma}+F_{a}^{\beta\gamma\alpha}+F_{a}^{\gamma\alpha\beta}), \end{equation}
where subindex $a$ equal $B$ for the BEC and $n$ for the normal fluid.

The fourth rank kinematic tensor $R_{a}^{\alpha\beta\gamma\delta}$ (\ref{BECTP20 R alpha beta gamma delta def})
has the following form after the introduction of the velocity field via the Madelung transformation of the many-particle wave function:
$$R_{a}^{\alpha\beta\gamma\delta}= n_{a}v_{a}^{\alpha}v_{a}^{\beta}v_{a}^{\gamma}v_{a}^{\delta}$$
$$+v_{a}^{\alpha}v_{a}^{\delta} (p_{a}^{\beta\gamma}+T_{a}^{\beta\gamma})
+v_{a}^{\beta}v_{a}^{\delta} (p_{a}^{\alpha\gamma}+T_{a}^{\alpha\gamma})
+v_{a}^{\gamma}v_{a}^{\delta} (p_{a}^{\alpha\beta}+T_{a}^{\alpha\beta})$$
$$+v_{a}^{\alpha}v_{a}^{\gamma} (p_{a}^{\beta\delta}+T_{a}^{\beta\delta})
+v_{a}^{\beta}v_{a}^{\gamma} (p_{a}^{\alpha\delta}+T_{a}^{\alpha\delta})
+v_{a}^{\alpha}v_{a}^{\beta} (p_{a}^{\gamma\delta}+T_{a}^{\gamma\delta})$$
$$+v_{a}^{\alpha}Q_{a}^{\beta\gamma\delta}
+v_{a}^{\beta}Q_{a}^{\alpha\gamma\delta}
+v_{a}^{\gamma}Q_{a}^{\alpha\beta\delta}
+v_{a}^{\delta}Q_{a}^{\alpha\beta\gamma}$$
\begin{equation} \label{BECTP20 R via p T Q}
+Q_{a}^{\alpha\beta\gamma\delta}+T_{a}^{\alpha\beta\gamma\delta}
+L_{a}^{\alpha\beta\gamma\delta}. \end{equation}
This structure shows some similarity to the representations for the second rank tensor momentum flux (\ref{BECTP20 Pi via n v p T})
and for the third rank tensor (\ref{BECTP20 M via p T Q}),
where the higher rank tensors are partially transformed via the concentration, velocity field and, if possible, via tensors of smaller rank.
However, this transformation is partial
since there is the tensor of the equal rank, but defined in the comoving frame.
Moreover, this final tensor is splitted on few parts.
It is two parts for the second rank tensor momentum flux,
where we have the kinetic pressure (quasi-classical part of thermal nature)
and the quantum Bohm potential (the quantum part).
There are three parts for the third rank tensor $M^{\alpha\beta\gamma}$.
They are the quasi-classical part of thermal nature, the quantum part, and the combined thermal-quantum part.
For the fourth rank tensor we also have three parts:
the quasi-classical part of thermal nature $Q_{a}^{\alpha\beta\gamma\delta}$,
the quantum part $T_{a}^{\alpha\beta\gamma\delta}$,
and the combined thermal-quantum part $L_{a}^{\alpha\beta\gamma\delta}$.

Developed model shows that arbitrary quantum system can be modeled via the hydrodynamic equations
which are traditionally associated with the fluid dynamics.
Quantum systems demonstrates that each particle shows the properties of the wave
and this wave-like behavior is incorporated in the quantum hydrodynamic model.
This conclusion follows from the fact that the quantum hydrodynamic is derived from the Schrodinger equations
which contains these information.
These general concept is illustrated for the ultracold bosons,
but the quantum hydrodynamic method can be applied to other physical systems.
This similarity between quantum behavior and the dynamics of fluids recently found unusual realization.
It is experimentally found that classic fluid objects demonstrate the quantum-like behavior \cite{Bush Ch 18},  \cite{Couder Nat 05}, \cite{Bush ARFM 15}.
It is observed as the millimetric droplet walking on the surface of vibrating fluids,
where the motion of droplets is affected by the resonant interaction with their own wave field \cite{Couder Nat 05}, \cite{Bush ARFM 15}.
Systems walking droplets demonstrate various quantum effects \cite{Cristea-Platon Ch 18}, \cite{Chowdury Ch 18}, \cite{Budanur Ch 19}.

\section{Contribution of interaction in the quantum hydrodynamic equations}

Equations (\ref{BECTP20 Euler eq 1 via j}), (\ref{BECTP20 eq for Pi alpha beta}),
and (\ref{BECTP20 eq for M alpha beta gamma})
contain terms describing interaction.
Approximate forms of these force fields of different tensor ranks are necessary to get a truncated set of equations.
In our case, it is necessary to include the short-range nature of the potential of the interparticle interaction.
Moreover, the weak interaction limit is considered.
These two assumptions are used to get simplified form of $F^{\alpha}$, $F^{\alpha\beta}$, $F^{\alpha\beta\gamma}$ and $F_{qf}^{\alpha\beta\gamma}$
in this section.

\subsection{Interaction terms in the Euler equation}

The short-range interaction in the Euler for the single species of quantum particles can be written as the divergence of the symmetric quantum stress tensor
$F^{\alpha}=-\partial^{\beta}\sigma^{\alpha\beta}$.

The first order by the interaction radius approximation gives the following expression for the quantum stress tensor
(see also \cite{Andreev PRA08})
\begin{equation}\label{BECTP20 sigma via psi}\sigma^{\alpha\beta}(\textbf{r},t)=-\frac{1}{2}\int
dR\sum_{i,j.i\neq j}\delta(\textbf{r}-\textbf{R}_{ij})\times$$
$$\times\frac{r^{\alpha}_{ij}r^{\beta}_{ij}}{\mid\textbf{r}_{ij}\mid}
\frac{\partial U(\textbf{r}_{ij})}{\partial\mid\textbf{r}_{ij}\mid}\Psi^{*}(R',t)\Psi(R',t), \end{equation}
where
$R'=\{..., \textbf{R}_{ij}, ..., \textbf{R}_{ij}, ...\}$
with vector $\textbf{R}_{ij}$ located at $i$-th and $j$-th places.

Expression (\ref{BECTP20 sigma via psi}) can be rewritten in terms of two-particle concentration
\begin{equation}\label{BECTP20 sigma via n2}
\sigma^{\alpha\beta}(\textbf{r},t)=-\frac{1}{2}Tr(n_{2}(\textbf{r},\textbf{r}',t))\int
d \textbf{r}\frac{r^{\alpha}r^{\beta}}{r}\frac{\partial U(r)}{\partial r},\end{equation}
where the notion of trace is used
\begin{equation}\label{BECTP20 Tr def} Tr f(\textbf{r},\textbf{r}')=f(\textbf{r},\textbf{r}).\end{equation}

Consideration of the short-range interaction leads to the separation of integral containing the potential of interaction.
So, the characteristic of interaction does not depend on the motion or position of particles.
This integral simplifies in the following way
\begin{equation}\label{BECTP20 int constant simplification}
\int\frac{r^{\alpha}r^{\beta}}{r}\frac{\partial U}{\partial r}d\textbf{r}
=\frac{1}{3}\delta^{\alpha\beta}\int rU' d\textbf{r}
=-\delta^{\alpha\beta}\int U d\textbf{r}.
\end{equation}
The last integral in this expression is denoted as
$g=\int U d\textbf{r}$.

The two-particle concentration can be calculated in the weakly interacting limit (see \cite{Andreev PRA08})
\begin{equation}\label{BECTP20 n2 long r}
n_2(\textbf{r},\textbf{r}',t)=n(\textbf{r},t)n(\textbf{r}',t)+|\rho(\textbf{r},\textbf{r}',t)|^{2}+\wp(\textbf{r},\textbf{r}',t)
,\end{equation}
where
\begin{equation}\label{BECTP20 nvarphi}
n(\textbf{r},t)=\sum_{f}n_{f}\varphi_{f}^{*}(\textbf{r},t)\varphi_{f}(\textbf{r},t)
\end{equation}
is the expression of concentration (\ref{BECTP20 concentration def b}) in terms of the single particle wave functions $\varphi_{f}(\textbf{r},t)$,
\begin{equation}\label{BECTP20 rhovarphi} \rho(\textbf{r}',\textbf{r},t)=\sum_{f}n_{f}\varphi_{f}^{*}(\textbf{r},t)\varphi_{f}(\textbf{r}',t)\end{equation}
is the density matrix,
and
\begin{equation}\label{BECTP20 def of DD function}
\wp(\textbf{r},\textbf{r}',t)=\sum_{f}n_{f}(n_{f}-1)|\varphi_{f}(\textbf{r},t)|^{2}|\varphi_{f}(\textbf{r}',t)|^{2},
\end{equation}
The last term in equation describes interaction of pairs of particles being in the same quantum state.
It can be seen from the existence of single quantum number $g$ in all wave are single-particle wave functions.

Expression (\ref{BECTP20 n2 long r}) can be substituted in the general expression of the force field (\ref{BECTP20 F alpha def via n2}).
However, equation (\ref{BECTP20 F alpha def via n2}) does not contain information about the short-range nature of considered interaction.
The first and second terms are related to particles located in different quantum states.
It cannot be seen from equation (\ref{BECTP20 n2 long r}),
but it follows from intermediate terms
which can be found in Ref. \cite{Andreev PRA08}.

The trace of the two-particle concentration entering the quantum stress tensor has the following form
\begin{equation}\label{BECTP20 n2 Tr via n}
Tr n_2(\textbf{r},\textbf{r}',t)\approx 2(n^{2})'+n^{2}_{B},\end{equation}
where the first term on the right-hand side symbol ' means that the product of concentrations is related to the particles in different quantum states.
Therefore, the first term has no $n^{2}_{B}$ contribution from selfaction of BEC.
The dropped terms are described in Ref. \cite{Andreev IJMP B 13}.

Present explicit contribution of the BEC concentration $n_{B}$
and the concentration of normal fluid $n_{n}$ in the first term on the right-hand side of equation (\ref{BECTP20 n2 Tr via n}):
\begin{equation}\label{BECTP20 n sqr '} (n^{2})'=((n_{B}+n_{n})(n_{B}+n_{n}))'=(n_{n}^{2}+ 2n_{B}n_{n}). \end{equation}

The last term in equation (\ref{BECTP20 n2 Tr via n}) appears for particles being in BEC.
While the first term on the right-hand side in equation (\ref{BECTP20 n2 Tr via n}) related to interaction of particles being in different quantum states.
Hence, it gives contribution for the interaction between BEC and normal fluid
\textit{and} for the interaction between bosons belonging to normal fluid.

Full expression of the quantum stress tensor for the bosons at finite temperature can be written
in terms of the concentration of BEC and the concentration of normal fluid:
\begin{equation}\label{BECTP20 sigma fin full}
\sigma^{\alpha\beta}=\frac{1}{2}g\delta^{\alpha\beta}(2n_{n}^{2}+4n_{B}n_{n}+n^{2}_{B}).\end{equation}

If we consider dynamics of BEC or normal fluid we cannot use the notion of the quantum stress tensor $\sigma^{\alpha\beta}$
for the interaction of subspecies
as it is for the interaction of different species.

The first (last) term in equation (\ref{BECTP20 sigma fin full}) contains the selfaction of the normal fluid (of the BEC).
The second term in equation (\ref{BECTP20 sigma fin full}) presents the interaction between the BEC and normal fluid.

If we consider dynamics of BEC we need to extract force caused by the BEC and normal fluid acting on the BEC.
This force is the superposition of  a part of the second term in equation (\ref{BECTP20 sigma fin full})
and the last term in equation (\ref{BECTP20 sigma fin full}):
\begin{equation}\label{BECTP20 sigma to F fin B}
F_{B}^{\alpha}=-g n_{B}\partial^{\alpha}(2n_{n}+n_{B}).\end{equation}
The second term in equation (\ref{BECTP20 sigma fin full})
can be rewritten as follows
$F_{2}^{\alpha}=-2g (n_{B}\partial^{\alpha}n_{n}+n_{n}\partial^{\alpha}n_{B})$.
The first part of this expression is used in equation (\ref{BECTP20 sigma to F fin B}).

If we consider dynamics of normal fluid
it means that the source of field in the first term of $n_2$ can be the normal fluid and the BEC,
hence the last term gives no contribution in this case in equation (\ref{BECTP20 sigma fin full}):
\begin{equation}\label{BECTP20 sigma to F fin n}
F_{n}^{\alpha}=-2g n_{n}\partial^{\alpha}(n_{n}+n_{B}).\end{equation}

The nonsymmetric decomposition allows to use the notion of the NLSE.
It is necessary condition to have the GP equation at finite temperatures.
Moreover, the nonsymmetric form is traditionally used in literature \cite{Griffin PRB 96}.
Same chose is made at analysis $j_{2}^{\beta}$ below.

\subsubsection{Nonlinear Schrodinger equations}

Dropping the pressure of normal fluid and using the quantum Bohm potential in form (\ref{BECTP20 Bohm tensor single part})
we find a closed set of hydrodynamic equations.
Introducing the macroscopic wave function for both the BEC and the normal fluid for the potential velocity fields
as $\Phi_{a}=\sqrt{n_{a}}e^{\imath m\phi_{a}/\hbar}$,
where $\phi_{a}$ is the potential of the velocity field $\textbf{v}_{a}=-\nabla\phi_{a}$.
\begin{equation}\label{BECTP20 GP B} \imath\hbar\partial_{t}\Phi_{B}=\Biggl(-\frac{\hbar^{2}\nabla^{2}}{2m}+V_{ext}+g(n_{B}+2n_{n})\Biggr)\Phi_{B}
,\end{equation}
and
\begin{equation}\label{BECTP20 GP like nf} \imath\hbar\partial_{t}\Phi_{n}=\Biggl(-\frac{\hbar^{2}\nabla^{2}}{2m}+V_{ext}+2g(n_{B}+n_{n})\Biggr)\Phi_{n}.\end{equation}

The kinetic energy (the first term on the right-hand side of equations (\ref{BECTP20 GP B}) and (\ref{BECTP20 GP like nf}))
corresponds to the application of the noninteracting limit for the quantum Bohm potential for the BEC and for the normal fluid.

The pressure of the normal fluid is dropped in equation (\ref{BECTP20 GP like nf}).

Equations (\ref{BECTP20 GP B}), (\ref{BECTP20 GP like nf}) correspond to equations 127-129 given in Ref. \cite{Dalfovo RMP 99}
while there is a difference in the form of presentation.

Therefore, the account of the pressure evolution together with the pressure flux evolution gives the generalization of the model presented with
the nonlinear Schrodinger equations (\ref{BECTP20 GP B}), (\ref{BECTP20 GP like nf}).
Necessity of additional equations is demonstrated in Refs. \cite{Andreev 2005}, \cite{Andreev 2007}, \cite{Andreev 2009}
if one wants to include the quantum fluctuations.

\subsection{Interaction terms in the pressure evolution equation}

General form of the pressure evolution equation contains the interaction via
the force second rank tensor field.
Its main contribution is obtained in the first order by the interaction radius.
The result appears in the following form
$$F^{\alpha\beta}(\textbf{r},t)=
\frac{1}{8m^{2}}
\partial^{\gamma}\int
dR\sum_{i,j;i\neq j}\delta(\textbf{r}-\textbf{R}_{ij})\times$$
$$\times\frac{r^{\beta}_{ij}r^{\gamma}_{ij}}{\mid\textbf{r}_{ij}\mid}\frac{\partial U(\textbf{r}_{ij})}{\partial\mid\textbf{r}_{ij}\mid} \biggl[\Psi^{*}(R',t)(\hat{p}_{(1)}^{\alpha}+\hat{p}_{(2)}^{\alpha})\Psi(R',t)+c.c.\biggr]$$
$$-\frac{1}{8m^{2}}
\int
dR\sum_{i,j;i\neq j}\delta(\textbf{r}-\textbf{R}_{ij})\frac{r^{\alpha}_{ij}r^{\gamma}_{ij}}{\mid\textbf{r}_{ij}\mid}\frac{\partial U(\textbf{r}_{ij})}{\partial\mid\textbf{r}_{ij}\mid}\times$$
$$\times \biggl[(\partial^{\gamma}_{(1)}-\partial^{\gamma}_{(2)})\Psi^{*}(R',t)(\hat{p}_{(1)}^{\alpha}-\hat{p}_{(2)}^{\alpha})\Psi(R',t) $$
\begin{equation}\label{BECTP20 F alpha beta via Psi FOIR} +\Psi^{*}(R',t)(\partial^{\gamma}_{(1)}-\partial^{\gamma}_{(2)})(\hat{p}_{(1)}^{\alpha}-\hat{p}_{(2)}^{\alpha})\Psi(R',t)+c.c.\biggr].
\end{equation}

Form (\ref{BECTP20 F alpha beta via Psi FOIR}) appears at the expansion of the force tensor field (\ref{BECTP20 F alpha beta def})
using the short-range nature of interaction
(see \cite{Andreev PRA08} for the method described for the force field,
or \cite{Andreev 2001} for application of this method to fermions).

For the force tensor field $F^{\alpha\beta}$
we can present the intermediate expressions like equations (\ref{BECTP20 sigma via n2}) and (\ref{BECTP20 n2 long r})
obtained for the force field $F^{\alpha}=-\partial^{\beta}\sigma^{\alpha\beta}$.
However, similar expressions obtained for $F^{\alpha\beta}$ are rather large.
Hence, we start the presentation with equation similar to equation (\ref{BECTP20 n2 Tr via n}) obtained after taking trace of the intermediate expressions.

Therefore, we obtain the following simplification of equation (\ref{BECTP20 F alpha beta via Psi FOIR})
for the force tensor field $F^{\alpha\beta}$:
$$F^{\alpha\beta}=-\frac{g}{4m^{2}}\partial^{\beta}[2(n\Lambda^{\alpha})'+ n_{B}\Lambda^{\alpha}_{B}]$$
\begin{equation}\label{BECTP20} +\frac{\imath}{\hbar}\frac{g}{4m^{2}}
[2(nr^{\alpha\beta})'-2(\Lambda^{\alpha}\Lambda^{\beta})'
+n_{B}r^{\alpha\beta}_{B}-\Lambda^{\alpha}_{B}\Lambda^{\beta}_{B}]
+c.c., \end{equation}
where
we use the intermediate functions $\Lambda^{\alpha}$ and $r^{\alpha\beta}$
with the following definitions:
\begin{equation}\label{BECTP20} \Lambda^{\alpha}=\sum_{f}n_{f}\varphi_{f}^{*}\hat{p}^{\alpha}\varphi_{f}
=mj^{\alpha}-\imath\frac{\hbar}{2}\partial^{\alpha}n, \end{equation}
and
$$r^{\alpha\beta}=\sum_{f}n_{f}\varphi_{f}^{*}\hat{p}^{\alpha}\hat{p}^{\beta}\varphi_{f}$$
$$=m^{2}\biggl(nv^{\alpha}v^{\beta}+p^{\alpha\beta}
-\frac{\hbar^{2}}{m^{2}}\sum_{f}n_{f}a_{f}\partial^{\alpha}\partial^{\beta}a_{f}\biggr)$$
\begin{equation}\label{BECTP20} -\imath\frac{m\hbar}{2}[\partial^{\alpha}(nv^{\beta})+\partial^{\beta}(nv^{\alpha})]. \end{equation}
The calculation of functions $\Lambda^{\alpha}$ and $r^{\alpha\beta}$
includes the Madelung transformation of the single-particle wave functions
$\varphi_{f}(r,t)=\sqrt{a_{f}}e^{\imath S_{f}}$.
Next, we use the following definitions of the velocity field and the pressure tensor
in terms of the single-particle wave functions
$nv^{\alpha}=\sum_{f}n_{f}a_{f}^{2}(\hbar\partial^{\alpha}S_{f}/m)$,
and
$p^{\alpha\beta}=\sum_{f}n_{f}a_{f}^{2}u_{f}^{\alpha}u_{f}^{\beta}$,
where
$u_{f}^{\alpha}=(\hbar\partial^{\alpha}S_{f}/m)-v^{\alpha}$.

Let us represent terms like $(n\Lambda^{\alpha})'$ in the explicit form:
$$F^{\alpha\beta}=-\frac{g}{4m^{2}}\partial^{\beta}[2n_{n}\Lambda^{\alpha}_{n}$$
$$+2n_{n}\Lambda^{\alpha}_{B}+2n_{B}\Lambda^{\alpha}_{n}+ n_{B}\Lambda^{\alpha}_{B}]$$
$$+\frac{\imath}{\hbar}\frac{g}{4m^{2}}
[2n_{n}r^{\alpha\beta}_{n}-2\Lambda^{\alpha}_{n}\Lambda^{\beta}_{n}
+2n_{n}r^{\alpha\beta}_{B}-2\Lambda^{\alpha}_{n}\Lambda^{\beta}_{B}$$
\begin{equation}\label{BECTP20}
2n_{B}r^{\alpha\beta}_{n}-2\Lambda^{\alpha}_{B}\Lambda^{\beta}_{n}
+n_{B}r^{\alpha\beta}_{B}-\Lambda^{\alpha}_{B}\Lambda^{\beta}_{B}]
+c.c.. \end{equation}

Further calculation gives the representation of tensor $F^{\alpha\beta}$
in term of hydrodynamic functions:
$$F^{\alpha\beta}=-\frac{g}{2m}\partial^{\beta}[2n_{n}j_{n}^{\alpha} +2n_{n}j_{B}^{\alpha} +2n_{B}j_{n}^{\alpha} +n_{B}j_{B}^{\alpha}]$$
$$+\frac{g}{4m}\biggl[2n_{n}(\partial^{}j_{n}^{}+\partial^{}j_{n}^{})-2 j_{n}^{}\partial^{}n_{n}-2 j_{n}^{}\partial^{}n_{n}$$
$$ +2n_{n}(\partial^{\alpha}j_{B}^{\beta}+\partial^{\beta}j_{B}^{\alpha})-2 j_{B}^{\alpha}\partial^{\beta}n_{n}-2 j_{B}^{\beta}\partial^{\alpha}n_{n} $$
$$ +2n_{B}(\partial^{\alpha}j_{n}^{\beta}+\partial^{\beta}j_{n}^{\alpha})-2 j_{n}^{\alpha}\partial^{\beta}n_{B}-2 j_{n}^{\beta}\partial^{\alpha}n_{B} $$
\begin{equation}\label{BECTP20}
+n_{B}(\partial^{\alpha}j_{B}^{\beta}+\partial^{\beta}j_{B}^{\alpha})- j_{B}^{\alpha}\partial^{\beta}n_{B}- j_{B}^{\beta}\partial^{\alpha}n_{B}\biggr]. \end{equation}

The momentum flux evolution equation contains the symmetric combination of the force tensor fields $F^{\alpha\beta}$:
$$F^{\alpha\beta}+F^{\beta\alpha}=
-\frac{g}{m}\biggl[2(j_{n}^{\alpha}\partial^{\beta}n_{n} +j_{n}^{\beta}\partial^{\alpha}n_{n})$$
$$+2(j_{n}^{\alpha}\partial^{\beta}n_{B} +j_{n}^{\beta}\partial^{\alpha}n_{B})$$
\begin{equation}\label{BECTP20} +2(j_{B}^{\alpha}\partial^{\beta}n_{n} +j_{B}^{\beta}\partial^{\alpha}n_{n})
+j_{B}^{\alpha}\partial^{\beta}n_{B}+j_{B}^{\beta}\partial^{\alpha}n_{B}\biggr].\end{equation}


The zero temperature analysis demonstrates that there is nonzero pressure for the BECs,
caused by the quantum fluctuations entering the set of hydrodynamic equations via the evolution of the pressure flux
\cite{Andreev 2005}, \cite{Andreev 2007}, \cite{Andreev 2009}.
The pressure also exists for the normal fluid.
So, we make decomposition of the momentum flux evolution equation on two partial equations
for $\Pi^{\alpha\beta}_{n}$ and $\Pi^{\alpha\beta}_{B}$.
Formally this decomposition is presented with equation (\ref{BECTP20 eq for Pi alpha beta Partial}).
To complete this procedure we need to split the force tensor field
$F^{\alpha\beta}+F^{\beta\alpha}$
$=F_{B}^{\alpha\beta}+F_{B}^{\beta\alpha}$
$+F_{n}^{\alpha\beta}+F_{n}^{\beta\alpha}$,
where
$$F_{B}^{\alpha\beta}+F_{B}^{\beta\alpha}=-\frac{g}{m}\biggl[2(j_{B}^{\alpha}\partial^{\beta}n_{n} +j_{B}^{\beta}\partial^{\alpha}n_{n})$$
\begin{equation}\label{BECTP20 Sigma to F B fin}
+j_{B}^{\alpha}\partial^{\beta}n_{B}+j_{B}^{\beta}\partial^{\alpha}n_{B}
\biggr],\end{equation}
and
$$F_{n}^{\alpha\beta}+F_{n}^{\beta\alpha}=-\frac{g}{m}\biggl[2(j_{n}^{\alpha}\partial^{\beta}n_{n} +j_{n}^{\beta}\partial^{\alpha}n_{n})$$
\begin{equation}\label{BECTP20 Sigma to F n fin}
+2(j_{n}^{\alpha}\partial^{\beta}n_{B} +j_{n}^{\beta}\partial^{\alpha}n_{B})\biggr]
.\end{equation}

After extraction of the pressure tensor $p^{\alpha\beta}$ from the momentum flux evolution $\Pi^{\alpha\beta}$
we have extra contribution of the interaction in the pressure evolution equation in compare with
equations (\ref{BECTP20 eq for Pi alpha beta Partial}).
It contains the following contribution $F^{\alpha\beta}-v^{\beta}F^{\alpha}$.

Using equations (\ref{BECTP20 sigma to F fin B}), (\ref{BECTP20 sigma to F fin n}), (\ref{BECTP20 Sigma to F n fin}), (\ref{BECTP20 Sigma to F B fin})
find $F^{\alpha\beta}+F^{\alpha\beta}-v^{\alpha}F^{\beta}-v^{\beta}F^{\alpha}=0$ for the BECs and for the normal fluid.

A pressure evolution equation is used in \cite{Kavoulakis PRA 98}
for bosons above the critical temperature.
Equation 4 of Ref. \cite{Kavoulakis PRA 98} contains the force in the following form $n\textbf{v}\cdot\textbf{F}$
which generally differs from $F^{\alpha\beta}-v^{\beta}F^{\alpha}$ obtained above.

\subsection{The short-range interaction in the third rank tensor evolution equation}

The third rank tensor $M^{\alpha\beta\gamma}$ (\ref{BECTP20 eq for M alpha beta gamma}) evolution equation contains two kinds of the
third rank force tensors $F^{\alpha\beta\gamma}$ (\ref{BECTP20 F alpha beta gamma def})
and $F_{qf}^{\alpha\beta\gamma}$ (\ref{BECTP20 F alpha beta gamma qf def}).
Consider them separately.

\subsubsection{Quasi-classical third rank force tensor}

Tensor
$F_{qf}^{\alpha\beta\gamma}$ (\ref{BECTP20 F alpha beta gamma qf def})
is proportional to the Planck constant,
so it goes to zero in the classical limit.
The third rank force tensor
$F^{\alpha\beta\gamma}$ (\ref{BECTP20 F alpha beta gamma def})
is different,
it has nonzero limit in the quasiclassical regime.
However, we are interested in the value of tensor
$F^{\alpha\beta\gamma}$ (\ref{BECTP20 F alpha beta gamma def})
in one of quantum regimes for the degenerate bosons.

We calculate the third rank force tensor
$F^{\alpha\beta\gamma}$ (\ref{BECTP20 F alpha beta gamma def})
in the first order by the interaction radius appears in the following form
\begin{widetext}
$$F^{\alpha\beta\gamma}(\textbf{r},t)=\frac{1}{8m^{3}}
\partial^{\mu}\int
dR\sum_{i,j;i\neq j}\delta(\textbf{r}-\textbf{R}_{ij})
\frac{r^{\mu}_{ij}r^{\gamma}_{ij}}{\mid\textbf{r}_{ij}\mid}\frac{\partial U(\textbf{r}_{ij})}{\partial\mid\textbf{r}_{ij}\mid} \biggl[\Psi^{*}(R',t)\hat{p}_{(1)}^{\alpha}\hat{p}_{(1)}^{\beta}\Psi(R',t)
+\hat{p}_{(1)}^{\beta *}\Psi^{*}(R',t)\hat{p}_{(1)}^{\alpha}\Psi(R',t)
+c.c.\biggr]$$

$$-\frac{1}{8m^{3}}\int
dR\sum_{i,j;i\neq j}\delta(\textbf{r}-\textbf{R}_{ij})\frac{r^{\mu}_{ij}r^{\gamma}_{ij}}{\mid\textbf{r}_{ij}\mid}\frac{\partial U(\textbf{r}_{ij})}{\partial\mid\textbf{r}_{ij}\mid}
\biggl[(\partial^{\mu}_{(1)}-\partial^{\mu}_{(2)})\Psi^{*}(R',t) \hat{p}_{(1)}^{\alpha}\hat{p}_{(1)}^{\beta}\Psi(R',t) $$
\begin{equation}\label{BECTP20 F alpha beta gamma via Psi FOIR}
+\Psi^{*}(R',t)
(\partial^{\mu}_{(1)}-\partial^{\mu}_{(2)})
\hat{p}_{(1)}^{\alpha}\hat{p}_{(1)}^{\beta}\Psi(R',t)
+\hat{p}_{(1)}^{\beta *}(\partial^{\mu}_{(1)}-\partial^{\mu}_{(2)})\Psi^{*}(R',t)\hat{p}_{(1)}^{\alpha}\Psi(R',t)
+\hat{p}_{(1)}^{\beta *}\Psi^{*}(R',t)(\partial^{\mu}_{(1)}-\partial^{\mu}_{(2)})\hat{p}_{(1)}^{\alpha}\Psi(R',t) +c.c.\biggr].\end{equation}

Here, the part of expression for $F^{\alpha\beta\gamma}$ containing the interaction potential appears as the independent multiplier.
It has same form as the integral in the Euler equation (\ref{BECTP20 int constant simplification}).
Hence, tensor $F^{\alpha\beta\gamma}$ is proportional to the Groos-Pitaevskii interaction constant.

Further calculation in the weakly interacting limit,
following the method described in Ref. \cite{Andreev PRA08},
gives an intermediate representation of the third rank force tensor:
\begin{equation}\label{BECTP20 F alpha beta gamma via Psi FOIR single particle WF}
F^{\alpha\beta\gamma}(\textbf{r},t)=-\frac{g}{4m^{3}}\Biggl[\Pi_{B}^{\alpha\beta}\partial^{\gamma}n_{B}+\Pi^{\alpha\beta}\partial^{\gamma}n
+\biggl(n \sum_{f}n_{f}\partial^{\gamma}\varphi_{f}^{*}\hat{p}^{\alpha}\hat{p}^{\beta}\varphi_{f}
+\frac{\imath}{\hbar}\Lambda^{\gamma}r^{\alpha\beta}
-\frac{\imath}{\hbar}\Lambda^{\alpha *}\kappa^{\gamma\beta}
+\frac{\imath}{\hbar}\Lambda^{\beta} \kappa^{\alpha\gamma}
+c.c.\biggr)\Biggr],\end{equation}
where
\begin{equation}\label{BECTP20} \kappa^{\alpha\beta}=
\sum_{f}n_{f}p^{\alpha}\varphi_{f}^{*}\cdot\hat{p}^{\beta} \varphi_{f}.\end{equation}
Function $\kappa^{\alpha\beta}$ is the nonsymmetric tensor.
It has symmetric real part and the antisymmetric imaginary part:
$$\kappa^{\alpha\beta}=m^{2}\biggl(nv^{\alpha}v^{\beta}+p^{\alpha\beta}
+\frac{\hbar^{2}}{m^{2}}\sum_{f}n_{f}\partial^{\alpha}\cdot a_{f}\partial^{\beta}a_{f}\biggr)$$
\begin{equation}\label{BECTP20} -\frac{1}{2}\imath m\hbar [v^{\alpha}\partial^{\beta}n-v^{\beta}\partial^{\alpha}n
+\sum_{f}n_{f}a_{f}(u^{\alpha}\partial^{\beta}a_{f}-u^{\beta}\partial^{\alpha}a_{f})].\end{equation}

No specific notation is introduced for the third rank tensor
$\sum_{f}n_{f}\partial^{\gamma}\varphi_{f}^{*}\hat{p}^{\alpha}\hat{p}^{\beta}\varphi_{f}$.
In our calculations we need its imaginary part multiplied by 2:
$$\sum_{f}n_{f}\partial^{\gamma}\varphi_{f}^{*}\hat{p}^{\alpha}\hat{p}^{\beta}\varphi_{f}+c.c.
=2m^{2}\Biggl[\frac{1}{2}\partial^{\gamma}n\cdot v^{\alpha}v^{\beta}
-\partial^{\alpha}n\cdot v^{\beta}v^{\gamma}
-\partial^{\beta}n\cdot v^{\alpha}v^{\gamma}
-nv^{\gamma}(\partial^{\beta}v^{\alpha}+\partial^{\alpha}v^{\beta})$$
$$+\sum_{f}n_{f}a_{f}(\partial^{\gamma}a_{f})u_{f}^{\alpha}u_{f}^{\beta}
-\frac{1}{2}\partial^{\beta}p^{\alpha\gamma}-\frac{1}{2}\partial^{\alpha}p^{\beta\gamma}
+\frac{1}{2}\sum_{f}n_{f}a_{f}^{2}(u_{f}^{\beta}\partial^{\alpha}u_{f}^{\gamma}+u_{f}^{\alpha}\partial^{\beta}u_{f}^{\gamma})$$
\begin{equation}\label{BECTP20}
+v^{\alpha}\sum_{f}n_{f}a_{f}(\partial^{\gamma}a_{f}\cdot u_{f}^{\beta}-\partial^{\beta}a_{f}\cdot u_{f}^{\gamma})
+v^{\beta}\sum_{f}n_{f}a_{f}(\partial^{\gamma}a_{f}\cdot u_{f}^{\alpha}-\partial^{\alpha}a_{f}\cdot u_{f}^{\gamma})
-\frac{\hbar^{2}}{m^{2}}\sum_{f}n_{f}\partial^{\gamma}a_{f}\cdot\partial^{\alpha}\partial^{\beta}a_{f}\biggr)\Biggr].\end{equation}

Equation (\ref{BECTP20 F alpha beta gamma via Psi FOIR single particle WF}) includes term $\Pi_{B}^{\alpha\beta}\partial^{\gamma}n_{B}$
which describes full contribution of the BEC in $F^{\alpha\beta\gamma}$.

The second term in equation (\ref{BECTP20 F alpha beta gamma via Psi FOIR single particle WF})
is an analog of the first term in equation (\ref{BECTP20 n2 long r}).
All following terms in equation (\ref{BECTP20 F alpha beta gamma via Psi FOIR single particle WF})
are the analog of the second term in equation (\ref{BECTP20 n2 long r}).
It can be interpreted as the exchange interaction.

Further calculation of the
(\ref{BECTP20 F alpha beta gamma via Psi FOIR single particle WF})
gives the partially truncated expression mainly presented via the macroscopic hydrodynamic functions:
$$F^{\alpha\beta\gamma}(\textbf{r},t)=-\frac{g}{4}\biggl[4\Pi_{B}^{\alpha\beta}\partial^{\gamma}n_{B}+4\Pi^{\alpha\beta}\partial^{\gamma}n
+\partial^{\alpha}n\biggl(\Pi^{\beta\gamma}+\frac{\hbar^{2}}{4m^{2}}\partial^{\beta}\partial^{\gamma}n\biggr)
+\partial^{\beta}n\biggl(\Pi^{\alpha\gamma}+\frac{\hbar^{2}}{4m^{2}}\partial^{\alpha}\partial^{\gamma}n\biggr)
+\partial^{\gamma}n\biggl(\Pi^{\alpha\beta}-\frac{\hbar^{2}}{4m^{2}}\partial^{\alpha}\partial^{\beta}n\biggr)$$
$$+n\biggl(3\partial^{\gamma}n\cdot v^{\alpha}v^{\beta}
-2\partial^{\alpha}n\cdot v^{\beta}v^{\gamma}
-2\partial^{\beta}n\cdot v^{\alpha}v^{\gamma}
-nv^{\gamma}(\partial^{\alpha}v^{\beta}+\partial^{\beta}v^{\alpha})
-\partial^{\beta}p^{\alpha\gamma}-\partial^{\alpha}p^{\beta\gamma}$$
$$+\sum_{f}n_{f}(\partial^{\gamma}a_{f}^{2})u_{f}^{\alpha}u_{f}^{\beta}
+\sum_{f}n_{f}a_{f}^{2}(u_{f}^{\beta}\partial^{\alpha}u_{f}^{\gamma}+u_{f}^{\alpha}\partial^{\beta}u_{f}^{\gamma})
+\frac{3}{2}v^{\alpha}\sum_{f}n_{f}(\partial^{\gamma}a_{f}^{2}\cdot u_{f}^{\beta}-\partial^{\beta}a_{f}^{2}\cdot u_{f}^{\gamma})$$
\begin{equation}\label{BECTP20 F alpha beta gamma fin}
+\frac{3}{2}v^{\beta}\sum_{f}n_{f}(\partial^{\gamma}a_{f}^{2}\cdot u_{f}^{\alpha}-\partial^{\alpha}a_{f}^{2}\cdot u_{f}^{\gamma})
-2\frac{\hbar^{2}}{m^{2}}\sum_{f}n_{f}\partial^{\gamma}a_{f}\cdot\partial^{\alpha}\partial^{\beta}a_{f}\biggr)\biggr]. \end{equation}

Equation for the third rank tensor $M^{\alpha\beta\gamma}$ evolution contains
the symmetric combination of the third rank force tensors (\ref{BECTP20 F alpha beta gamma fin})
which cam be presented in the following form:
$$F^{\alpha\beta\gamma}+F^{\beta\gamma\alpha}+F^{\gamma\alpha\beta}
=-\frac{g}{4}
\biggl[4(\Pi_{B}^{\beta\gamma}\partial^{\alpha}n_{B}+\Pi_{B}^{\alpha\gamma}\partial^{\beta}n_{B}+\Pi_{B}^{\alpha\beta}\partial^{\gamma}n_{B})
+4(\Pi^{\beta\gamma}\partial^{\alpha}n+\Pi^{\alpha\gamma}\partial^{\beta}n+\Pi^{\alpha\beta}\partial^{\gamma}n)$$
$$+\partial^{\alpha}n\biggl(3\Pi^{\beta\gamma}+\frac{\hbar^{2}}{4m^{2}}\partial^{\beta}\partial^{\gamma}n\biggr)
+\partial^{\beta}n\biggl(3\Pi^{\alpha\gamma}+\frac{\hbar^{2}}{4m^{2}}\partial^{\alpha}\partial^{\gamma}n\biggr)
+\partial^{\gamma}n\biggl(3\Pi^{\alpha\beta}+\frac{\hbar^{2}}{4m^{2}}\partial^{\alpha}\partial^{\beta}n\biggr)$$
\begin{equation}\label{BECTP20 F alpha beta gamma symm}
-n(\partial^{\alpha}\Pi^{\beta\gamma}+\partial^{\beta}\Pi^{\alpha\gamma}+\partial^{\gamma}\Pi^{\alpha\beta})
-\frac{3\hbar^{2}}{4m^{2}}n\partial^{\alpha}\partial^{\beta}\partial^{\gamma}n\biggr].\end{equation}

The pressure flux evolution equation obtained as the reduction of the third rank tensor $M^{\alpha\beta\gamma}$ evolution equation
contains the following combination of the force fields:
$$F^{\alpha\beta\gamma}+F^{\beta\gamma\alpha}+F^{\gamma\alpha\beta}
-\frac{1}{mn}(F^{\alpha}\Pi^{\beta\gamma}+F^{\beta}\Pi^{\alpha\gamma}+F^{\gamma}\Pi^{\alpha\beta})
=
\frac{g}{4}[\partial^{\alpha}(n\Pi^{\beta\gamma})+\partial^{\beta}(n\Pi^{\alpha\gamma})+\partial^{\gamma}(n\Pi^{\alpha\beta})]'$$
\begin{equation}\label{BECTP20 F alpha beta gamma symm reduced} +\frac{g\hbar^{2}}{16m^{2}}[3n\partial^{\alpha}\partial^{\beta}\partial^{\gamma}n
-\partial^{\alpha}n\cdot\partial^{\beta}\partial^{\gamma}n
-\partial^{\beta}n\cdot\partial^{\alpha}\partial^{\gamma}n
-\partial^{\gamma}n\cdot\partial^{\alpha}\partial^{\beta}n]', \end{equation}
where symbol $[]'$ specifies that product of functions describing the BEC is excluded
similarly equations
(\ref{BECTP20 n2 Tr via n}) and (\ref{BECTP20 n sqr '}).

The quantum part can be represented
$$F^{\alpha\beta\gamma}+F^{\beta\gamma\alpha}+F^{\gamma\alpha\beta}
-\frac{1}{mn}(F^{\alpha}\Pi^{\beta\gamma}+F^{\beta}\Pi^{\alpha\gamma}+F^{\gamma}\Pi^{\alpha\beta})
=
\frac{g}{4}[\partial^{\alpha}(n\Pi^{\beta\gamma})+\partial^{\beta}(n\Pi^{\alpha\gamma})+\partial^{\gamma}(n\Pi^{\alpha\beta})]'$$
\begin{equation}\label{BECTP20 F alpha beta gamma symm reduced II}
+\frac{g\hbar^{2}}{16m^{2}}[\partial^{\alpha}(n\partial^{\beta}\partial^{\gamma}n-\partial^{\beta}n\cdot\partial^{\gamma}n)
+\partial^{\beta}(n\partial^{\alpha}\partial^{\gamma}n-\partial^{\alpha}n\cdot\partial^{\gamma}n)
+\partial^{\gamma}(n\partial^{\alpha}\partial^{\beta}n-\partial^{\alpha}n\cdot\partial^{\beta}n)]'. \end{equation}

It can be considered as
$F^{\alpha\beta\gamma}+F^{\beta\gamma\alpha}+F^{\gamma\alpha\beta}
-\frac{1}{mn}(F^{\alpha}\Pi^{\beta\gamma}+F^{\beta}\Pi^{\alpha\gamma}+F^{\gamma}\Pi^{\alpha\beta})$
$=\tilde{F}^{\alpha\beta\gamma}+\tilde{F}^{\beta\gamma\alpha}+\tilde{F}^{\gamma\alpha\beta}$,
where
\begin{equation}\label{BECTP20 F alpha beta gamma symm reduced II}
\tilde{F}^{\alpha\beta\gamma}=\frac{g}{4}[\partial^{\alpha}(n\Pi^{\beta\gamma})
+\frac{g\hbar^{2}}{16m^{2}}[\partial^{\alpha}(n\partial^{\beta}\partial^{\gamma}n-\partial^{\beta}n\cdot\partial^{\gamma}n)]'.
\end{equation}
It is the derivative of the second rank tensor.

If we consider the zero temperature limit
we find $F^{\alpha\beta\gamma}=(-g/4m^{3})\Pi_{B}^{\alpha\beta}\partial^{\gamma}n_{B}$.
The transition from the equation of evolution for tensor $M^{\alpha\beta\gamma}$
to the equation of evolution for tensor $Q^{\alpha\beta\gamma}$,
which is the sibling of $M^{\alpha\beta\gamma}$, but the pressure flux $Q^{\alpha\beta\gamma}$ is defined in the comoving frame
leads to the canceling of such term.
Hence, the nonzero contribution comes from the quantum part of the third rank force tensor $F_{qf}^{\alpha\beta\gamma}$.
The nonzero temperature gives a nonzero contribution of $F^{\alpha\beta\gamma}$ at the transition to the pressure flux evolution equation.

Equation (\ref{BECTP20 F alpha beta gamma symm reduced II}) is obtained for all bosons.
We need to separate it on the force acting on the BEC and the force acting on the normal fluid.

Finally, we obtain
$$\tilde{F}_{n}^{\alpha\beta\gamma}+\tilde{F}_{n}^{\beta\gamma\alpha}+\tilde{F}_{n}^{\gamma\alpha\beta}
=
\frac{g}{4}[\partial^{\alpha}(n_{n}\Pi_{n}^{\beta\gamma})+\partial^{\beta}(n_{n}\Pi_{n}^{\alpha\gamma})+\partial^{\gamma}(n_{n}\Pi_{n}^{\alpha\beta})$$
$$+n_{n}\partial^{\alpha}(\Pi_{B}^{\beta\gamma}+\partial^{\beta}\Pi_{B}^{\alpha\gamma}+\partial^{\gamma}\Pi_{B}^{\alpha\beta})
+\Pi_{n}^{\beta\gamma}\partial^{\alpha}n_{B}+\Pi_{n}^{\alpha\gamma}\partial^{\beta}n_{B}+\Pi_{n}^{\alpha\beta}\partial^{\gamma}n_{B}]$$
$$+\frac{g\hbar^{2}}{16m^{2}}\biggl[3n_{n}\partial^{\alpha}\partial^{\beta}\partial^{\gamma}n_{n}
+3n_{n}\partial^{\alpha}\partial^{\beta}\partial^{\gamma}n_{B}
-\partial^{\alpha}n_{n}\cdot\partial^{\beta}\partial^{\gamma}n_{n}
-\partial^{\beta}n_{n}\cdot\partial^{\alpha}\partial^{\gamma}n_{n}
-\partial^{\gamma}n_{n}\cdot\partial^{\alpha}\partial^{\beta}n_{n}$$
\begin{equation}\label{BECTP20 F alpha beta gamma symm reduced BEC}
-\frac{1}{2}\partial^{\alpha}n_{n}\cdot\partial^{\beta}\partial^{\gamma}n_{B}
-\frac{1}{2}\partial^{\beta}n_{n}\cdot\partial^{\alpha}\partial^{\gamma}n_{B}
-\frac{1}{2}\partial^{\gamma}n_{n}\cdot\partial^{\alpha}\partial^{\beta}n_{B}
-\frac{1}{2}\partial^{\alpha}n_{B}\cdot\partial^{\beta}\partial^{\gamma}n_{n}
-\frac{1}{2}\partial^{\beta}n_{B}\cdot\partial^{\alpha}\partial^{\gamma}n_{n}
-\frac{1}{2}\partial^{\gamma}n_{B}\cdot\partial^{\alpha}\partial^{\beta}n_{n}\biggr], \end{equation}
and
$$\tilde{F}_{B}^{\alpha\beta\gamma}+\tilde{F}_{B}^{\beta\gamma\alpha}+\tilde{F}_{B}^{\gamma\alpha\beta}
=
\frac{g}{4}[
n_{B}\partial^{\alpha}(\Pi_{n}^{\beta\gamma}+\partial^{\beta}\Pi_{n}^{\alpha\gamma}+\partial^{\gamma}\Pi_{n}^{\alpha\beta})$$
$$+\Pi_{B}^{\beta\gamma}\partial^{\alpha}n_{n}+\Pi_{B}^{\alpha\gamma}\partial^{\beta}n_{n}+\Pi_{B}^{\alpha\beta}\partial^{\gamma}n_{n}]
+\frac{g\hbar^{2}}{16m^{2}}\biggl[3n_{B}\partial^{\alpha}\partial^{\beta}\partial^{\gamma}n_{n}$$
\begin{equation}\label{BECTP20 F alpha beta gamma symm reduced nf}
-\frac{1}{2}\partial^{\alpha}n_{B}\cdot\partial^{\beta}\partial^{\gamma}n_{n}
-\frac{1}{2}\partial^{\beta}n_{B}\cdot\partial^{\alpha}\partial^{\gamma}n_{n}
-\frac{1}{2}\partial^{\gamma}n_{B}\cdot\partial^{\alpha}\partial^{\beta}n_{n}
-\frac{1}{2}\partial^{\alpha}n_{n}\cdot\partial^{\beta}\partial^{\gamma}n_{B}
-\frac{1}{2}\partial^{\beta}n_{n}\cdot\partial^{\alpha}\partial^{\gamma}n_{B}
-\frac{1}{2}\partial^{\gamma}n_{n}\cdot\partial^{\alpha}\partial^{\beta}n_{B}\biggr]. \end{equation}
Equations (\ref{BECTP20 F alpha beta gamma symm reduced BEC}) and (\ref{BECTP20 F alpha beta gamma symm reduced nf}) gives
final expressions for the quasi-classic force fields in the pressure flux evolution equations for two fluid model.

\end{widetext}

\subsubsection{The third rank force tensor describing the quantum fluctuation}

The quantum fluctuations in the zero temperature BECs is caused by tensor
$F_{qf}^{\alpha\beta\gamma}$ (\ref{BECTP20 F alpha beta gamma qf def}).
Its major contribution can be found in the first order by the interaction radius approximation.
Here, we consider the small nonzero temperature regime of $F_{qf}^{\alpha\beta\gamma}$ for the bosons.
So, we obtain its generalization for the two-fluid model.
The quantum third rank force tensor $F_{qf}^{\alpha\beta\gamma}$ (\ref{BECTP20 F alpha beta gamma qf def})
is calculated in the first order by the interaction radius
$$F_{qf}^{\alpha\beta\gamma}=-\frac{\hbar^{2}}{8m^{2}}
\partial^{\delta}
\int dR\sum_{i,j.i\neq j}\delta(\textbf{r}-\textbf{R}_{ij})\times$$
\begin{equation} \label{BECTP20 F alpha beta gamma qf TOIR}
\times r^{\delta}_{ij}\partial^{\alpha}_{i}\partial^{\beta}_{i}\partial^{\gamma}_{i} U(\textbf{r}_{ij})
\Psi^{*}(R',t)\Psi(R',t). \end{equation}
In formula (\ref{BECTP20 F alpha beta gamma qf TOIR}) for tensor $F_{qf}^{\alpha\beta\gamma}$
we have separation of the integral containing the interaction potential,
as we have at the calculation of other force fields above.
However, here we obtain different integral $\int r^{\alpha}\partial^{\beta}\partial^{\gamma}\partial^{\delta}$.
Calculation of this integral leads to the second interaction constant given below
in the simplified expression for the quantum third rank force tensor
\begin{equation} \label{BECTP20 F alpha beta gamma qf TOIR via n2}
F_{qf}^{\alpha\beta\gamma}
=\frac{\hbar^{2}}{8m^{2}} g_{2}I_{0}^{\alpha\beta\gamma\delta}\partial^{\delta}
Tr n_{2}(\textbf{r},\textbf{r}',t), \end{equation}
where
\begin{equation} \label{BECTP20 def g 2} g_{2}=\frac{2}{3}\int d\textbf{r} U''(r), \end{equation}
and
\begin{equation} \label{BECTP20 I 4} I_{0}^{\alpha\beta\gamma\delta}=\delta^{\alpha\beta}\delta^{\gamma\delta} +\delta^{\alpha\gamma}\delta^{\beta\delta}+\delta^{\alpha\delta}\delta^{\beta\gamma}.  \end{equation}

Calculation of the two-particle concentration leads to the following expression:
\begin{equation} \label{BECTP20 F alpha beta gamma qf TOIR via n with nonzero T}
F_{qf}^{\alpha\beta\gamma}
=\frac{\hbar^{2}}{8m^{2}} g_{2}I_{0}^{\alpha\beta\gamma\delta}
\partial^{\delta}(2n_{n}^{2}+4n_{B}n_{n}+n_{B}^{2}).
\end{equation}
Here we have $\partial^{\delta}(2n_{n}^{2}+4n_{B}n_{n}+n_{B}^{2})$.
Next, we open brackets and find
$(4n_{n}\partial^{\delta}n_{n}+4n_{n}\partial^{\delta}n_{B}+4n_{B}\partial^{\delta}n_{n}+2n_{B}\partial^{\delta}n_{B})$.
The source of field is under derivative, while the multiplier in front corresponds to the species under the action of field.
Hence the first two terms correspond to the force acting on the normal fluid,
while the third and fourth terms correspond to the force acting on the BEC.

Therefore, we can separate expression (\ref{BECTP20 F alpha beta gamma qf TOIR via n with nonzero T})
on two parts corresponding to the BEC and to the normal fluid:
\begin{equation} \label{BECTP20 F alpha beta gamma qf TOIR via n with nonzero T for BEC}
F_{qf,B}^{\alpha\beta\gamma}
=\frac{\hbar^{2}}{4m^{2}} g_{2}I_{0}^{\alpha\beta\gamma\delta}(2n_{B}\partial^{\delta}n_{n}+n_{B}\partial^{\delta}n_{B}),
\end{equation}
and
\begin{equation} \label{BECTP20 F alpha beta gamma qf TOIR via n with nonzero T for nf}
F_{qf,n}^{\alpha\beta\gamma}
=\frac{\hbar^{2}}{2m^{2}} g_{2}I_{0}^{\alpha\beta\gamma\delta}\partial^{\delta}(n_{n}\partial^{\delta}n_{n}+n_{B}\partial^{\delta}n_{n}).
\end{equation}


\section{Hydrodynamic equations for two fluid model of bosons with nonzero temperature}

This section provides the final set of equations obtained in this paper.
The method of the introduction of the velocity field and corresponding representation of the hydrodynamic equations is not described in this paper.
It can be found in number other papers, majority of details are given in Refs. \cite{Andreev PRA08}, \cite{Andreev 2001}.

In this regime we have two continuity equations:
\begin{equation}\label{BECTP20 cont eq bosons BEC} \partial_{t}n_{B}+\nabla\cdot (n_{B}\textbf{v}_{B})=0, \end{equation}
and
\begin{equation}\label{BECTP20 cont eq bosons excited} \partial_{t}n_{n}+\nabla\cdot (n_{n}\textbf{v}_{n})=0. \end{equation}

The Euler equation for bosons in the BEC state
$$mn_{B}(\partial_{t} +\textbf{v}_{B}\cdot\nabla)v^{\alpha}_{B}
+\partial_{\beta}T_{B}^{\alpha\beta}+\partial_{\beta}p_{qf}^{\alpha\beta}$$
\begin{equation}\label{BECTP20 Euler bosons BEC}
+g n_{B}\partial^{\alpha}n_{B} =-n_{B}\partial^{\alpha}V_{ext}
-2g n_{B}\partial^{\alpha}n_{n},\end{equation}
where the quantum Bohm potential is given by equation (\ref{BECTP20 Bohm tensor single part}) in the noninteracting limit.

The Euler equation for bosons in the excited states corresponding to the nonzero temperature
$$mn_{n}(\partial_{t} +\textbf{v}_{n}\cdot\nabla)v^{\alpha}_{n}+\partial_{\beta}p_{n,eff}^{\alpha\beta}$$
\begin{equation}\label{BECTP20 Euler bosons excited}
+2g n_{n}\partial^{\alpha}n_{n} =-n_{n}\partial^{\alpha}V_{ext}
-2g n_{n}\partial^{\alpha}n_{B},\end{equation}
where the effective pressure tensor $p_{n,eff}^{\alpha\beta}=p_{n}^{\alpha\beta}+T_{n}^{\alpha\beta}$.

The effective pressure evolution equation for normal boson fluid is also a part of developed and applied hydrodynamic model
$$\partial_{t}p_{n,eff}^{\alpha\beta}
+v_{n}^{\gamma}\partial_{\gamma}p_{n,eff}^{\alpha\beta}
+p_{n,eff}^{\alpha\gamma}\partial_{\gamma}v_{n}^{\beta}
+p_{n,eff}^{\beta\gamma}\partial_{\gamma}v_{n}^{\alpha}$$
\begin{equation} \label{BECTP20 pressure evolution n}
+p_{n,eff}^{\alpha\beta}\partial_{\gamma}v_{n}^{\gamma}
+\partial_{\gamma}T^{\alpha\beta\gamma}_{n}
+\partial_{\gamma}Q^{\alpha\beta\gamma}_{n}
=0.  \end{equation}

Moreover, we have the pressure (the quantum Bohm potential)
$p_{B,eff}^{\alpha\beta}=T_{B}^{\alpha\beta}+p_{qf}^{\alpha\beta}$ evolution equation for the BEC
$$\partial_{t}p_{B,eff}^{\alpha\beta} +v_{B}^{\gamma}\partial_{\gamma}p_{B,eff}^{\alpha\beta}
+p_{B,eff}^{\alpha\gamma}\partial_{\gamma}v_{B}^{\beta} +p_{B,eff}^{\beta\gamma}\partial_{\gamma}v_{B}^{\alpha}$$
\begin{equation} \label{BECTP20 qBp evolution}
+p_{B,eff}^{\alpha\beta}\partial_{\gamma}v_{B}^{\gamma}+\partial_{\gamma}T^{\alpha\beta\gamma}_{B}
+\partial_{\gamma}Q^{\alpha\beta\gamma}_{qf}
=0.  \end{equation}

Let us to point out the following property of the quantum Bohm potential
that it satisfies the following equation for the arbitrary species $a$
$$\partial_{t}T_{a}^{\alpha\beta} +v_{a}^{\gamma}\partial_{\gamma}T_{a}^{\alpha\beta}
+T_{a}^{\alpha\gamma}\partial_{\gamma}v_{a}^{\beta} +T_{a}^{\beta\gamma}\partial_{\gamma}v_{a}^{\alpha}$$
\begin{equation} \label{BECTP20 qBp evolution a}
+T_{a}^{\alpha\beta}\partial_{\gamma}v_{a}^{\gamma}
+\partial_{\gamma}T^{\alpha\beta\gamma}_{a}
=0. \end{equation}
It is expected that approximate form of the quantum Bohm potential (\ref{BECTP20 Bohm tensor single part}) satisfies
equation (\ref{BECTP20 qBp evolution a}) existing at the zero interaction.
Hence, substitute (\ref{BECTP20 Bohm tensor single part}) in equation (\ref{BECTP20 qBp evolution a}) with the zero right-hand side:
$$\partial^{\beta}\partial^{\gamma}n_{a}\cdot(\partial^{\gamma}v_{a}^{\alpha}-\partial^{\alpha}v_{a}^{\gamma})
+\partial^{\alpha}\partial^{\gamma}n_{a}\cdot(\partial^{\gamma}v_{a}^{\beta}-\partial^{\beta}v_{a}^{\gamma})$$
$$+\frac{1}{3}\partial_{\gamma}n_{a}\cdot (\partial^{\beta}\partial^{\gamma}v_{a}^{\alpha} +\partial^{\alpha}\partial^{\gamma}v_{a}^{\beta}-\partial^{\alpha}\partial^{\beta}v_{a}^{\gamma})$$
\begin{equation} \label{BECTP20 qBp in equation}
+\frac{1}{3}n_{a} [\triangle(\partial^{\beta}v_{a}^{\alpha} +\partial^{\alpha}v_{a}^{\beta})
-\partial^{\alpha}\partial^{\beta}(\nabla \cdot\textbf{v}_{a})]=0,\end{equation}
where the continuity equation is used for the time derivatives of concentration.
Make the condition of the potentiality of the velocity field $\textbf{v}_{a}=\nabla\phi_{a}$.
Include it in equation (\ref{BECTP20 qBp in equation}) and find that this equation is satisfied.

Hence, we obtain the simplified form of the pressure evolution equations,
where the traditional quantum Bohm potential is extracted:
$$\partial_{t}p_{qf,B}^{\alpha\beta}
+v_{B}^{\gamma}\partial_{\gamma}p_{qf,B}^{\alpha\beta}
+p_{qf,B}^{\alpha\gamma}\partial_{\gamma}v_{B}^{\beta}
+p_{qf,B}^{\beta\gamma}\partial_{\gamma}v_{B}^{\alpha}$$
\begin{equation} \label{BECTP20 pressure evolution n no T0}
+p_{qf,B}^{\alpha\beta}\partial_{\gamma}v_{B}^{\gamma}
+\partial_{\gamma}Q^{\alpha\beta\gamma}_{qf,B}
=0,  \end{equation}
and
$$\partial_{t}p_{n}^{\alpha\beta}
+v_{n}^{\gamma}\partial_{\gamma}p_{n}^{\alpha\beta}
+p_{n}^{\alpha\gamma}\partial_{\gamma}v_{n}^{\beta}
+p_{n}^{\beta\gamma}\partial_{\gamma}v_{n}^{\alpha}$$
\begin{equation} \label{BECTP20 pressure evolution n no T0}
+p_{n}^{\alpha\beta}\partial_{\gamma}v_{n}^{\gamma}
+\partial_{\gamma}Q^{\alpha\beta\gamma}_{n}
=0.  \end{equation}

Equation for the evolution of quantum-thermal part of the third rank tensor is \cite{Andreev 2005}, \cite{Andreev 2007}:
$$\partial_{t}Q_{qf}^{\alpha\beta\gamma} +\partial_{\delta}(v_{B}^{\delta}Q_{qf}^{\alpha\beta\gamma})
+Q_{qf}^{\alpha\gamma\delta}\partial_{\delta}v_{B}^{\beta}
+Q_{qf}^{\beta\gamma\delta}\partial_{\delta}v_{B}^{\alpha}
+Q_{qf}^{\alpha\beta\delta}\partial_{\delta}v_{B}^{\gamma}$$
$$=\frac{\hbar^{2}}{4m^{2}} g_{2}I_{0}^{\alpha\beta\gamma\delta}\biggl(n_{B}\partial^{\delta}n_{B}+2n_{B}\partial^{\delta}n_{n})\biggr)$$
$$-\frac{1}{m}n_{B}\partial_{\alpha}\partial_{\beta}\partial_{\gamma}V_{ext}
+\tilde{F}_{B}^{\alpha\beta\gamma}+\tilde{F}_{B}^{\beta\gamma\alpha}+\tilde{F}_{B}^{\gamma\alpha\beta}$$
\begin{equation} \label{BECdqfS20 eq evolution Q qf}
+\frac{1}{mn}(p_{qf,eff}^{\alpha\beta}\partial^{\delta}p_{qf,eff}^{\gamma\delta}
+p_{qf,eff}^{\alpha\gamma}\partial^{\delta}p_{qf,eff}^{\beta\delta}
+p_{qf,eff}^{\beta\gamma}\partial^{\delta}p_{qf,eff}^{\alpha\delta}),  \end{equation}
and
$$\partial_{t}Q_{n}^{\alpha\beta\gamma} +\partial_{\delta}(v_{n}^{\delta}Q_{n}^{\alpha\beta\gamma})
+Q_{n}^{\alpha\gamma\delta}\partial_{\delta}v_{n}^{\beta}
+Q_{n}^{\beta\gamma\delta}\partial_{\delta}v_{n}^{\alpha}
+Q_{n}^{\alpha\beta\delta}\partial_{\delta}v_{n}^{\gamma}$$
$$=\frac{\hbar^{2}}{2m^{2}} g_{2}I_{0}^{\alpha\beta\gamma\delta}\biggl(n_{n}\partial^{\delta}n_{n}+n_{n}\partial^{\delta}n_{B}\biggr)$$
$$-\frac{1}{m}n_{n}\partial_{\alpha}\partial_{\beta}\partial_{\gamma}V_{ext}
+\tilde{F}_{n}^{\alpha\beta\gamma}+\tilde{F}_{n}^{\beta\gamma\alpha}+\tilde{F}_{n}^{\gamma\alpha\beta}$$
\begin{equation} \label{BECdqfS20 eq evolution Q qf}
+\frac{1}{mn}(p_{n,eff}^{\alpha\beta}\partial^{\delta}p_{n,eff}^{\gamma\delta}
+p_{n,eff}^{\alpha\gamma}\partial^{\delta}p_{n,eff}^{\beta\delta}
+p_{n,eff}^{\beta\gamma}\partial^{\delta}p_{n,eff}^{\alpha\delta}),  \end{equation}
where
$\tilde{F}_{a}^{\alpha\beta\gamma}$ is not presented explicitly
since equations (\ref{BECTP20 F alpha beta gamma symm reduced BEC})
and (\ref{BECTP20 F alpha beta gamma symm reduced nf}) show that required expressions are rather large.

Refs. \cite{Andreev PRA08}, \cite{Andreev LP 19}.
Hydrodynamic model for fermions with pressure evolution is derived in
Refs. \cite{Andreev 2001}, \cite{Andreev 1912}, \cite{Andreev LP 21}.

Terms proportional to
$p_{a,eff}^{\alpha\beta}\partial^{\delta}p_{a,eff}^{\gamma\delta}$
appears in the pressure flux evolution equation,
but it leads to the contribution beyond the chosen approximation
\cite{Tokatly PRB 99}, \cite{Tokatly PRB 00}.

Term containing the external potential
$-\frac{1}{m}n_{a}\partial_{\alpha}\partial_{\beta}\partial_{\gamma}V_{ext}$
goes to zero for the parabolic trap.
However, it can give some nontrivial contribution for other form of potentials.

\section{BEC dynamics under the influence of the quantum fluctuations}

Developed model shows that
there is nontrivial evolution equation for the pressure and the pressure flux of the BEC.
Therefore, the well-known model of BEC is extended in spite the fact that
the kinetic pressure tensor is expected to be equal to zero due to zero temperature.
However, the quantum fluctuations lead to the nonzero occupation numbers for the excited states.

If we need to consider pure BEC
we need to drop the contribution of the normal fluid in the model presented above.
Therefore, let us summarize the BEC model in parabolic traps:
\begin{equation}\label{BECTP20 cont eq bosons BEC B}
\partial_{t}n_{B}+\nabla\cdot (n_{B}\textbf{v}_{B})=0, \end{equation}
$$mn_{B}(\partial_{t} +\textbf{v}_{B}\cdot\nabla)v^{\alpha}_{B}
+\partial_{\beta}(p_{qf}^{\alpha\beta}+T_{B}^{\alpha\beta})$$
\begin{equation}\label{BECTP20 Euler bosons BEC B}
+g n_{B}\partial^{\alpha}n_{B} +n_{B}\partial^{\alpha}V_{ext}=0
,\end{equation}
$$\partial_{t}p_{qf}^{\alpha\beta}
+v_{B}^{\gamma}\partial_{\gamma}p_{qf}^{\alpha\beta}
+p_{qf}^{\alpha\gamma}\partial_{\gamma}v_{B}^{\beta}
+p_{qf}^{\beta\gamma}\partial_{\gamma}v_{B}^{\alpha}$$
\begin{equation} \label{BECTP20 pressure evolution BEC T=0}
+p_{qf}^{\alpha\beta}\partial_{\gamma}v_{B}^{\gamma}
+\partial_{\gamma}Q^{\alpha\beta\gamma}_{qf}
=0,  \end{equation}
and
$$\partial_{t}Q_{qf}^{\alpha\beta\gamma} +\partial_{\delta}(v^{\delta}Q_{qf}^{\alpha\beta\gamma})
+Q_{qf}^{\alpha\gamma\delta}\partial_{\delta}v^{\beta}$$
\begin{equation} \label{BECdqfS20 eq evolution Q qf}
+Q_{qf}^{\beta\gamma\delta}\partial_{\delta}v^{\alpha}
+Q_{qf}^{\alpha\beta\delta}\partial_{\delta}v^{\gamma}
=\frac{\hbar^{2}}{4m^{2}} g_{2}I_{0}^{\alpha\beta\gamma\delta}n\partial^{\delta}n.  \end{equation}
This simplified model is reported earlier in
Refs. \cite{Andreev 2005}, \cite{Andreev 2007}, \cite{Andreev 2009},
where the dipole-dipole interaction is also considered.
It has been demonstrated that the quantum fluctuations gives mechanisms for the instability of the small amplitude perturbations \cite{Andreev 2005}.
Moreover, the dipolar part of the quantum fluctuations creates conditions for the bright soliton in the repulsive BECs \cite{Andreev 2009}.
The developed in previous section model provides proper generalization of earlier model giving the small temperature contribution.

\section{Conclusion}

Revision of the two-fluid model for the finite temperature ultracold bosons has been presented
through the derivation of the number of particles, the momentum, the momentum flux, and the third rank tensor balance equations.
The derivation has been based on the trace of the microscopic dynamics of quantum particles
via the application of the many-particle microscopic Schrodinger equation.
Hence, the microscopic Schrodinger equation determines the time evolution
for the macroscopic functions describing the collective motion of bosons.

General equations have been derived for the arbitrary strength of interaction and the arbitrary temperatures.
The set of equations has been restricted by the third rank kinematic tensor (the flux of pressure).
The truncation is made for the low temperatures weakly interacting bosons
after the derivation of the general structure of hydrodynamic equations.
Therefore, the thermal part of the fourth rank kinematic tensor has been taken equal to zero.
Next, the terms containing the interaction potential have been considered for the short-range interaction.
The small radius of interaction provides the small parameter for the expansion.
The expansion is made in the force field in the Euler equation,
the force tensor field in the momentum flux equation,
and the third rank force tensor in the pressure flux evolution equation.
The first term in expansion on the small interparticle distance has been considered in each expansion,
which corresponds to the first order by the interaction radius.

This model allows to gain the quantum fluctuations
which is the essential property of the BECs.
Moreover, the interaction causing the quantum fluctuations has been consider at the finite temperature.

The functions obtained in the first order by the interaction radius have been expressed via the trace of two-particle functions.
The two-particle functions have been calculated for the weakly interacting bosons.

The single species of 0-spin bosons has been considered.
Therefore, the single fluid hydrodynamics has been derived.
Next, it has been included that the concentration of particles, the current of particles (the momentum density),
the momentum flux,
and the current of the momentum flux are additive functions.
Consequently, they can be easily splitted on two parts: the BEC and the normal fluid of bosons (the non BEC part).
Hence, the two fluid model of single species of bosons is obtained.
This separation on two fluids has been made in general form of equations.

\section{Acknowledgements}

Work is supported by the Russian Foundation for Basic Research (grant no. 20-02-00476).
This paper has been supported by the RUDN University Strategic Academic Leadership Program.

\end{document}